\documentclass[twocolumn]{article}
\usepackage{graphicx}

\def\cref#1{Chapt.\,\ref{#1}}
\def\Cref#1{Chapter~\ref{#1}}
\def\sref#1{Sect.\,\ref{#1}}

\def\fref#1{Fig.\,\ref{#1}}

\def\rref#1{Ref.\,\cite{#1}}

\def\Cerenkov{\v{C}erenkov}
\def\deg{$^\circ$}

\def\Section#1{\section{#1}}
\def\iref#1#2{\{{\sc #1} \#{\it #2}\}}

\newcommand{\etal}{\MakeLowercase{\textit{et al.}}} 
\begin{document}

{\bf Highlights in astroparticle physics: muons, neutrinos, hadronic
interactions, exotic particles, and dark matter --- Rapporteur Talk HE2 \& HE3\\[0.5\baselineskip]}


{J.R.~H\"orandel\\[0.5\baselineskip]}

{Department of Astrophysics, IMAPP, Radboud University Nijmegen, 
 6500 GL Nijmegen, The Netherlands --- http://particle.astro.ru.nl\\[0.5\baselineskip]}


{\sl
Recent results presented at the International Cosmic Ray Conference in Beijing
will be reviewed. Topics include HE2: "Muons and Neutrinos" and HE3:
"Interactions, Particle Physics Aspects, Cosmology".\\[0.5\baselineskip]
}

{\bf keywords:} {muons, neutrinos, neutrino telescopes, hadronic interactions,
          exotic particles, dark matter}

\section{Introduction}
The 32nd International Cosmic Ray Conference was held in August 2011 in
Beijing.
About 150 papers were presented in the sessions
HE2: "Muons and Neutrinos" 
and
HE3: "Interactions, Particle Physics Aspects, Cosmology".
Some of the highlights presented at the conference will be reviewed.

A wide spectrum of (astro-)physical questions is addressed in the papers
covered by this rapporteur. The common theme seems to be "upper limits and
bounds".
The topics include
\begin{itemize}
\item Underground experiments, measuring the muon flux from air showers, find
  evidence for anisotropies in the arrival direction of cosmic rays with TeV
  energies (\sref{he21}).
\item The sensitivity of neutrino oscillation searches with water/ice
  \Cerenkov\ detectors has been reported. \\
   Supernova searches indicate that
   the rate of gravitational stellar collapses in the Galaxy is less than 0.13
   events/year (\sref{he22}).
\item The big neutrino telescopes ANTARES and IceCube are fully operational.
  Limits on the diffuse neutrino flux reach sensitivities around
  $10^{-8}$~GeV~cm$^{-2}$ s$^{-1}$ sr$^{-1}$. No point sources have (yet) 
  been found (\sref{he23}).
\item A new method has been introduced, the "endpoint formalism", to calculate
   electromagnetic radiation from any kind of particle acceleration
   (\sref{he24}).
\item Muons are used in a geophysical application for tomography of a volcano
  (\sref{he25}).
\item New projects are underway to detect high-energy neutrinos: KM3NeT in the
  Mediterranean Sea, as well as ARA and ARIANNA on the Antarctic continent
  (\sref{he26}).

\item First data from the Large Hadron Collider (LHC), in particular from the
  forward detector LHCf, yield new insight into hadronic interactions, which are
  of great importance to describe the development of extensive air showers
  (\sref{he31}).
\item New upper limits on magnetic monopoles reach sensitivities of the order of
  $10^{-18}$~cm$^{-2}$~s$^{-1}$~sr$^{-1}$. Searches for antinuclei indicate
  that there is less than 1 antihelium nucleus per $10^7$ helium nuclei in the
  Universe (\sref{he33}). 
\item Astrophysical dark matter searches yield upper limits for the
  velocity-weighted annihilation cross section of the order of 
  $\langle\sigma v\rangle<10^{-24}$~cm$^3$~s$^{-1}$ (\sref{he34}).
\end{itemize}


\Section{HE 2.1 Muon experiments}\label{he21}
\subsection{Cosmic-ray anisotropy}
The muons registered with GeV energies in underground muon detectors are
produced in air showers with TeV energies. Thus, studying the arrival
directions of muons in underground facilities provides a mean to study the
arrival directions of TeV primary cosmic rays.

At present, the biggest underground muon detector is IceCube at the South Pole.
It is comprised of a 1~km$^3$ water/ice \Cerenkov\ detector at a depth below
1450~m w.e.\ \cite{arxiv1105-2326} \iref{BenZvi}{306}.  \footnote{References in
\{\} refer to the Proceedings of the 32nd International Cosmic Ray Conference,
Beijing (2011) and to the slides available on the conference web site
http://icrc2011.ihep.ac.cn/ \{{\it (first) author name} \#{\it paper
number}\}.} The detector is composed of 80 strings, each equipped with 60
optical sensor modules. A surface array (IceTop) consists of 81 stations,
each composed of 2 \Cerenkov\ detector water ice tanks, read out by two optical
sensors, respectively. 

\begin{figure}[t]
 \includegraphics[width=\columnwidth]{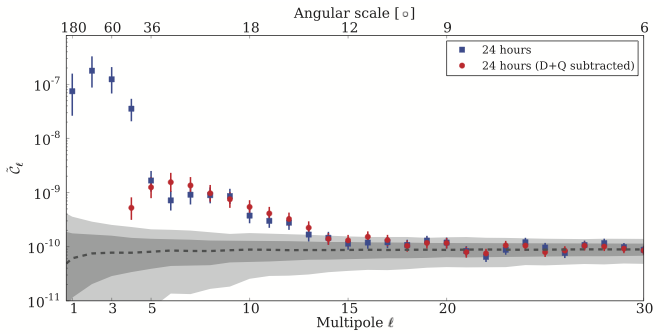}
 \caption{Angular power spectra for the cosmic ray intensity observed by 
          IceCube, see also \fref{icecube-sky} \cite{arxiv1105-2326} 
          \iref{BenZvi}{306}.}
 \label{icecube-multipole}
\end{figure}

\begin{figure}[t]
 \includegraphics[width=\columnwidth]{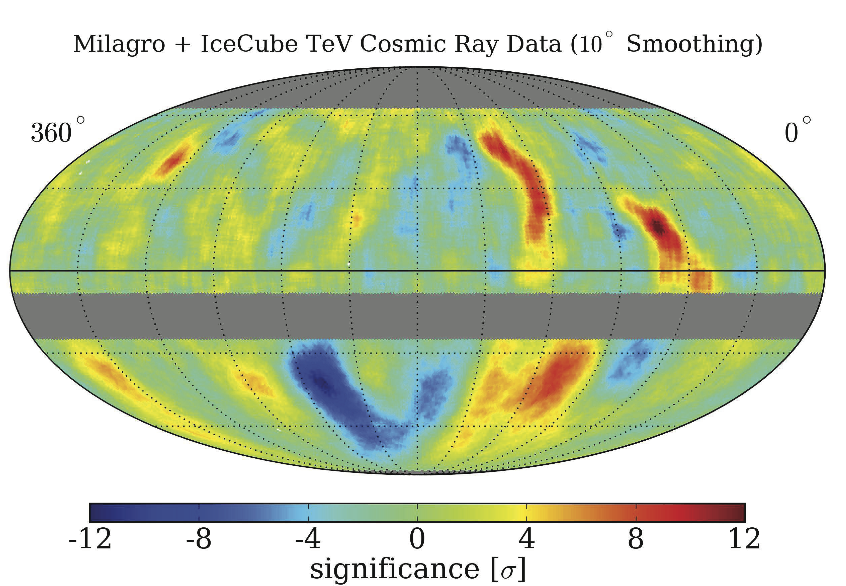}  
 \caption{Arrival directions of TeV cosmic rays observed by IceCube and
          Milagro \cite{arxiv1105-2326} \iref{BenZvi}{306}.}
 \label{icecube-sky}
\end{figure}

With the more than $32\cdot10^9$ events recorded, it is possible to probe the
southern sky for per-mille anisotropy on all angular scales in the arrival
direction distribution of cosmic rays. A power spectrum analysis of
the relative intensity map of the cosmic-ray flux reveals that the arrival
direction distribution is not isotropic, see \fref{icecube-multipole}.
Significant structures are seen on several angular scales: a large-scale
structure in the form of a strong dipole and a quadrupole, as well as
small-scale structure on scales between 15\deg\ and 30\deg.  The skymap exhibits
several localized regions of significant excess and deficit in cosmic-ray
intensity, see \fref{icecube-sky}. The IceCube observations complement
measurements from other detectors in the northern hemisphere, such as the
Milagro experiment. The origin of this anisotropy is still unknown.

Another detector in the northern hemisphere is the MINOS neutrino oscillation
experiment \iref{de Jong}{1185}.  It is comprised of a near and a far detector,
two steel-scintillator sampling calorimeters, installed at 100~m and 700~m
underground, respectively. Muon data from the near detector have been used to
investigate the arrival direction of cosmic rays.  A sky map of the
significances indicates anisotropies which are compatible with observations by
ARGO-YBJ. A projection of the arrival directions on the right ascension axis
yields a relative anisotropy amplitude on the order of 0.1\%.

\subsection{Annual modulation} 
Annual variations in the (average) temperature profile of the atmosphere alter
the air density and, thus, influence the development of air showers.  This
affects in particular the decay of pions into muons and yields to a seasonal
modulation of the observed muon flux.  

\begin{figure*}
 \includegraphics[width=\textwidth]{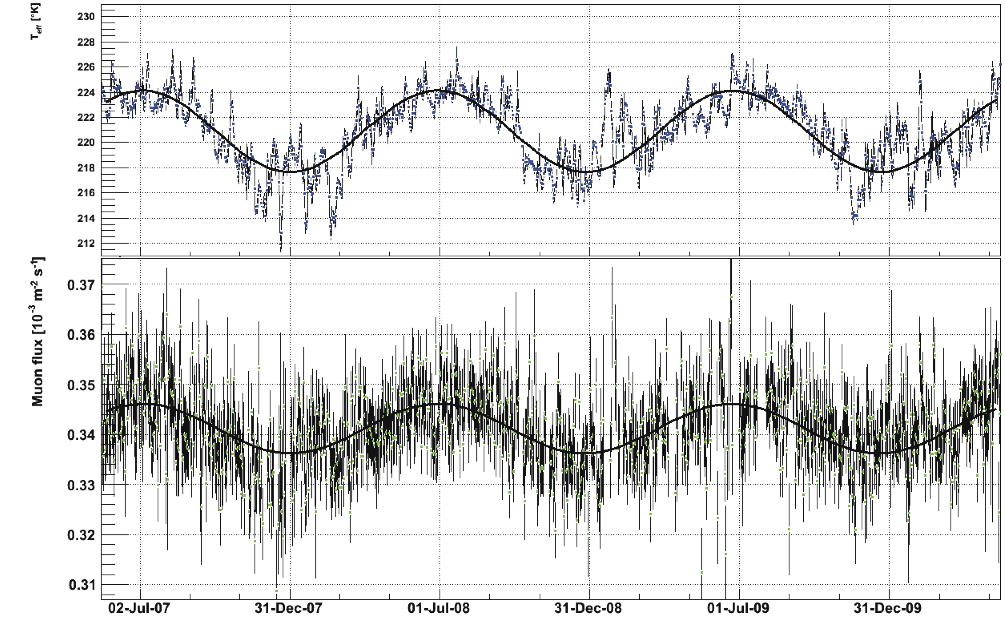}
 \caption{Cosmic-ray induced muon flux observed by BOREXINO as function of time
          \iref{d'Angelo}{510}.}
 \label{borexinoannual}
\end{figure*}
The BOREXINO solar neutrino detector is located at Gran Sasso National
Laboratory below a rock coverage corresponding to about 3800~m w.e.\
\iref{d'Angelo}{510}.  It is a liquid scintillation detector with an active
mass of about 1.33 kt.  The measured muon flux as function of time (see
\fref{borexinoannual}) exhibits a seasonal modulation with an amplitude of
$~1.5\%$.  The muon flux reaches its maximum at day $186.2\pm0.4$.

It is remarkable that the measured muon phase almost agrees with the phase of a
dark matter signal, claimed by the DAMA experiment at the same location
\cite{arxiv1007-595}. The maximum of the dark matter signal is observed on June
2nd (day 152).  This correlation has also been pointed out in a recent paper
\cite{Blum:2011jf}, using muon flux measurements from the LVD detector.  It
remains unclear if the seasonal modulation of the underground muon flux
significantly influences the observed dark matter signal, after taking into
account various measures to reduce the muon-related background.

\subsection{Solar modulation}
Solar modulation of Galactic cosmic rays is investigated with GRAND
\iref{Poirier}{1292}.  This is an array of 64 proportional wire chamber
stations dispersed in an area of 100~m $\times$ 100~m. It continuously monitors
the muon flux recorded for about 20 years.  The muon data extend the energy
range covered by the world-wide neutron monitor network to higher energies.  An
example is the Forbush decrease on October 29th, 2003, when GRAND registered a
sudden intensity drop of 8\%.  In addition, the direction of the incoming
muons is recorded, giving hints of the disturbances in the structure of the
heliosphere during a Forbush decrease.

\subsection{Muon interactions}
When a high-energy muon traverses matter, such as Antarctic ice, catastrophic
energy losses occasionally occur \iref{Berghaus}{85}. With IceCube the
longitudinal profile of such events is reconstructed. This provides a
calorimetric measurement of the cascades with TeV energies.  The technique is
used to extend the measured atmospheric muon spectrum to higher energies up to
500~TeV.

A similar method is applied in the ANTARES neutrino telescope to reconstruct
the energy of electromagnetic showers induced by muons in water
\iref{Mangano}{158}. The cascades are induced by TeV muons via pair production
and bremsstrahlung. Muons with energies up to 100~TeV have been reconstructed.

In extensive air showers muons with high transverse momentum $p_T$ are
occasionally produced \iref{Gerhardt}{323}. The dominant production processes
are the semi\-leptonic decay of heavy quarks and the decay of high $p_T$
kaons and pions in jets. High $p_T$ muons manifest themselves in the data as
single muons separated a few hundred meters from the shower core. IceCube with
its surface detector IceTop is well suited to study such effects. High $p_T$
muons can be reconstructed with a minimum resolvable $p_T=8$~GeV for a 1~TeV
muon.  A data sample of slightly more than 100 days, taken during the
construction phase of the detector has been analyzed. No high $p_T$ muon has
been found.

\subsection{Muon charge ratio}
Recent measurements of the atmospheric muon charge ratio for momenta around
1~GeV/c are in agreement with expectations from calculations
\iref{Brancus}{400} \iref{Abdolahi}{749}.

\Section{HE 2.2 Solar, atmospheric, and related neutrino experiments}
\label{he22}

\begin{figure*}
 \includegraphics[width=\textwidth]{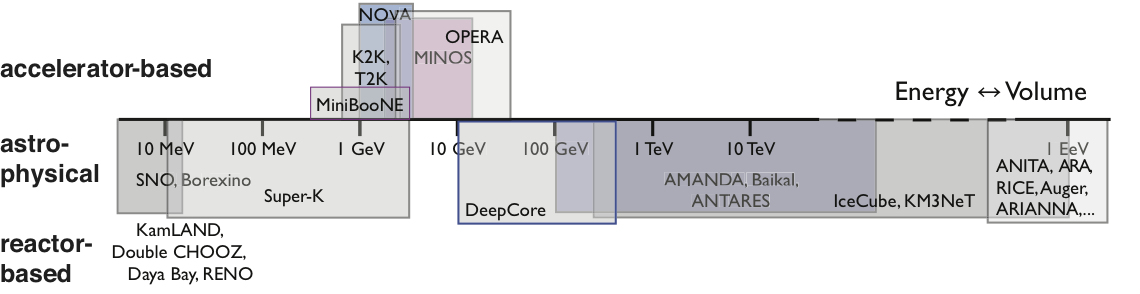}
 \caption{Energy range of various neutrino detectors, after \iref{Ha}{324}.}
 \label{nu-spec}
\end{figure*} 

In present experiments, neutrinos are measured over a wide energy range. An
overview on several detectors and their energy range is given in
\fref{nu-spec}.  Astrophysical neutrinos are expected from MeV energies (solar
neutrinos) up to the EeV range (presumably extragalactic sources).  The
properties of neutrinos are investigated using neutrinos generated in nuclear
reactors (MeV regime) and accelerators (GeV regime).
Recent results on neutrino oscillation measurements are discussed in
\sref{he22} and the detection of high-energy neutrinos in \sref{he23}.

Neutrino oscillations are commonly described in terms of the $L/E$ dependence,
where $E$ is the neutrino energy and $L$ its oscillation path length
\cite{Dore:2008dp}. For a neutrino telescope such as ANTARES or IceCube,
detecting neutrinos crossing the Earth, $L$ can be translated as $2R \sin
\Theta$, $R$ being the Earth radius and $\Theta$ the angle between the neutrino
direction and the vertical axis ($\Theta=\pi/2$ for a vertical upgoing
neutrino). Within the two-flavor approximation, the $\nu_\mu$ survival
probability can then be written as
$$P(\nu_\mu\rightarrow \nu_\mu) \approx 1-\sin^2 2\vartheta_{23}
   \sin^2\left( 2.54 R~\Delta m^2_{23} \frac{\sin\Theta}{E}\right).$$
$\vartheta_{23}$ and $\Delta m_{23}$ being respectively the mixing angle and
the squared mass difference of the involved mass eigenstates (with $R$ in km,
$E$ in GeV and $\Delta m_{23}$ in eV$^2$).  According to recent results from
the MINOS experiment \cite{Adamson:2011ig}, the first minimum in the muon
neutrino survival probability $E/\sin \Theta$ occurs at about 24~GeV.
This energy range can be reached with modern neutrino telescopes.

\subsection{ANTARES}
The ANTARES neutrino telescope is located in the Mediterranean Sea, 40 km south
of the French coast at a depth of about 2500~m below sea level.  The detector
is an array of photomultiplier tubes arranged on 12 vertical detector lines.
Each string is comprised of up to 25 floors, each composed of a triplet of
optical modules.

\begin{figure}[t]
 \includegraphics[width=\columnwidth]{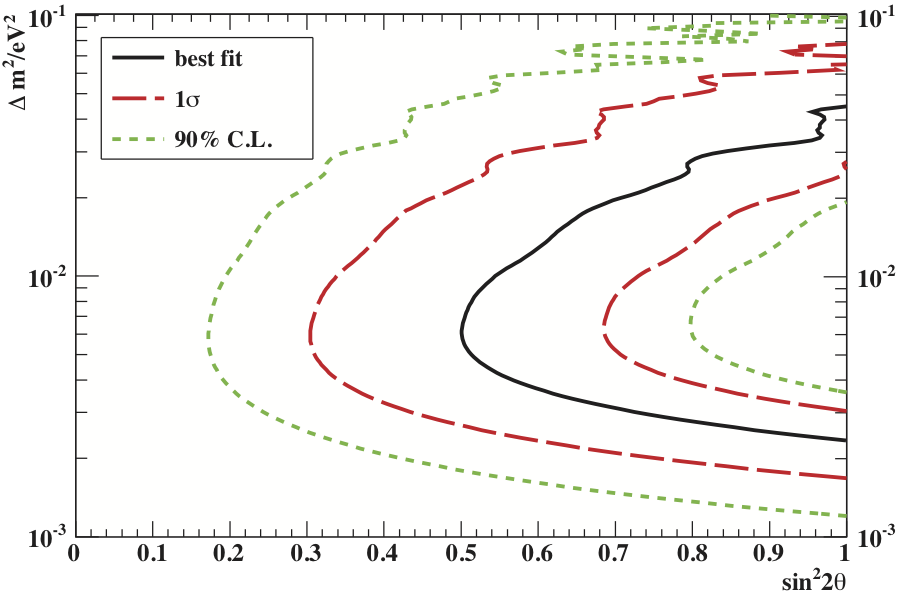}
 \caption{Expected sensitivity of ANTARES to atmospheric neutrino oscillation
          parameters after 1000 days measuring time \iref{Guillard}{819}.}
 \label{antares-sens}
\end{figure} 
The ANTARES collaboration plans to search for neutrino oscillations
\iref{Guillard}{819}.  Studies have shown that a clean sample of atmospheric
neutrinos with energies as low as 20~GeV can be isolated with the neutrino
telescope.  The expected sensitivity after 1000 days, corresponding to a total
measuring time of four to five years of data-taking, is shown in
\fref{antares-sens}.  The number of expected events is roughly one per day,
which for 1000 days leads to a statistical error of about 3\% ($\approx30$
events).  According to these results, although not competitive with dedicated
experiments, ANTARES should be sensitive to neutrino oscillation parameters
through the disappearance of atmospheric muon neutrinos.

\subsection{IceCube}
The IceCube detector has an extension to measure neutrinos with lower energies,
IceCube DeepCore \iref{Ha}{324}. This are eight strings in the center of the
array which contain additional optical sensors at the bottom of the instrument
with a spacing optimized for lower energies. 

IceCube DeepCore can study atmospheric neutrino oscillations through a
combination of its low energy reach, as low as about 10 GeV, and its
unprecedented statistical sample, of about 150000 triggered atmospheric muon
neutrinos per year \iref{Euler}{329}. 
As described above, the muon neutrino disappearance minimum and tau neutrino
appearance maximum are expected at about 25 GeV, which is considerably lower in
energy than typical IceCube neutrino events, but higher than the energies at
which accelerator-based experiments have detected oscillations.  $\nu_\mu$
disappearance and $\nu_\tau$ appearance from neutrino oscillations can be
measured in IceCube.  Simulations indicate that for certain oscillation
parameters, the measured $\nu_\mu$ flux is expected to differ by about a factor
of two from the flux expected without oscillations.

IceCube is also suited to serve as a supernova detector by monitoring the
optical module counting rates across the array \iref{Baum}{1137}.  It is
expected to detect subtle features in the temporal development of MeV neutrinos
from the core collapse of nearby massive stars. For a supernova at the Galactic
center, its sensitivity matches that of a background-free megaton-scale
supernova search experiment and is expected to trigger on supernovae with about
200, 20, and 6 standard deviations at the Galactic center (10 kpc), the
Galactic edge (30 kpc), and the Large Magellanic Cloud (50 kpc), respectively.

\subsection{Super-Kamiokande}
The Super-Kamiokande detector, one of the experiments establishing neutrino
oscillations \cite{Fukuda:1998mi,Fukuda:2001nk}, has recently been upgraded
with new electronics \iref{Carminati}{723} and is now capable of detecting
thermal neutrons from neutrino interactions \iref{Zhang}{353}.  The objective
is to identify low-energy anti-neutrinos via inverse beta decay reactions and
to reduce the atmospheric neutrino background for nucleon decay searches.  A
delayed-coincidence detection of a positron and a neutron capture offers a
powerful way to identify low-energy anti-neutrinos via the inverse beta decay
reaction
$\bar{\nu}_e+\mbox{p}\rightarrow \mbox{e}^+ + \mbox{n}$.
The subsequent neutron capture in water yields a photon pair:
$\mbox{n}+\mbox{p}\rightarrow d+\gamma$ (2.2~MeV).

\begin{figure}[t]
 \includegraphics[width=\columnwidth]{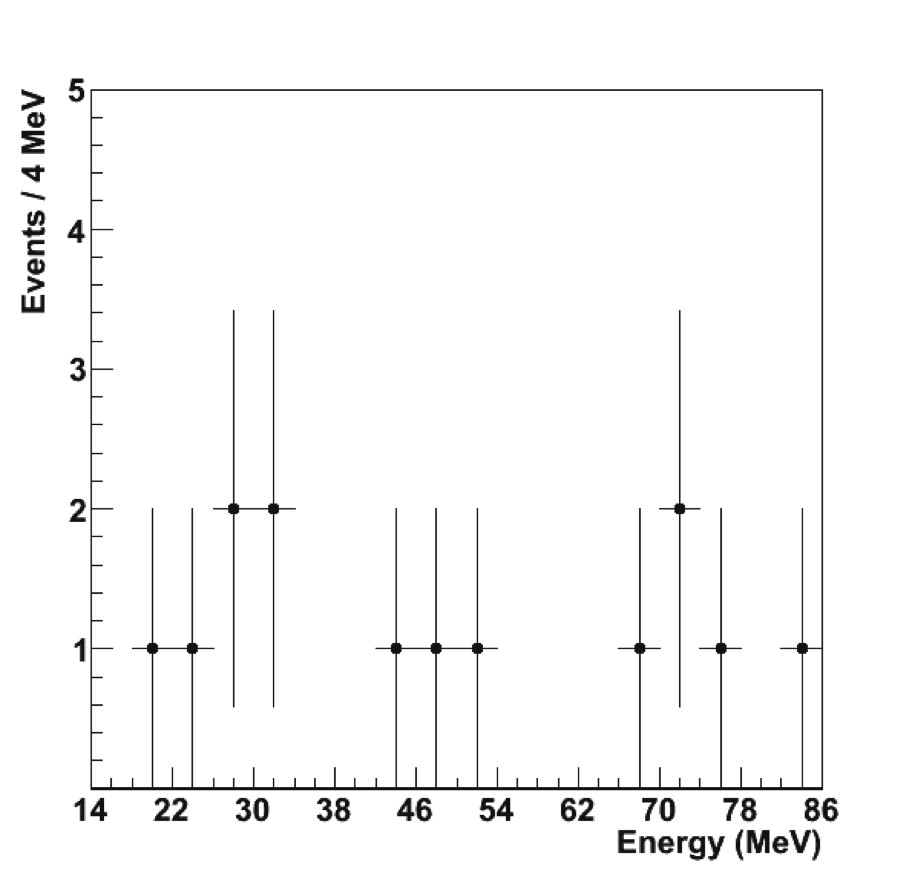}
 \caption{Neutrino energy spectrum ($E_\nu >14$~MeV) reconstructed with neutron
          tagging in Super-Kamiokande \iref{Zhang}{353}.}
 \label{superk}
\end{figure} 
The new technique has been verified by measurements which exhibit a clear
neutron signal in atmospheric neutrino data.  The neutrino energy spectrum has
been extended to lower energies ($E_\nu>14$~MeV), as depicted in \fref{superk}.

In searches for supernova relic neutrinos or GUT monopoles in Super-Kamiokande,
nuclear de-excitation $\gamma$-rays induced by neutral current interactions of
atmospheric neutrinos with $^{16}$O are one of the major background sources
\iref{Ueno}{253}.  Such events are studied with the T2K experiment, which
started in January 2010.  Its beam energy is typical to atmospheric neutrinos.
A low-energy ($4-30$~MeV) event search has been performed in the T2K data. This
yields a better understanding of the low-energy background.

The improved sensitivity is used to search for neutrinos from distant
supernovae \iref{Smy}{855}. No evidence of a signal has been found.  An upper
limit (90\% c.l.) on the neutrino flux has been established of about
2.7~neutrinos/cm$^2$/s above 17.3~MeV neutrino energy.

\subsection{ICARUS}
The ICARUS detector, located at the Gran Sasso Underground Laboratory is
composed of a big liquid argon time projection chamber (TPC), containing 600~t
of liquid argon \iref{Cocco}{1340}.  ICARUS is detecting neutrinos from the
CNGS beam, produced at CERN and reaching Gran Sasso after a flight of about
730~km under the Earth surface.  Data-taking has started in October 2010 and
the first neutrinos from the CNGS beam have been measured at Gran Sasso.
The time projection chamber exhibits excellent imaging capability together with
good spatial and calorimetric resolutions.
The main goal is to search for $\nu_\mu\rightarrow\nu_\tau$ oscillations in the
neutrino beam from CERN.
The expected sensitivity in the $\Delta$m$^2 -\sin^2(2\Theta)$ plot is expected
to fully cover the LNSD-allowed region below $\Delta\mbox{m}^2<1$~eV$^2$/c$^2$.

\subsection{LVD}
The Large Volume Detector (LVD), located at the Gran Sasso Underground
Laboratory, is a 1~kt liquid scintillator neutrino observatory, mainly designed
to study low-energy neutrinos from gravitational stellar collapses in the
Galaxy \iref{Fulgione}{513}.  The experiment has been taking data since June
1992, with increasingly larger mass configurations. The telescope duty cycle,
in the last ten years, has been greater than 99\%.  Recent results include the
search for neutrino bursts, analyzing data from May 1st, 2009 to March 27th,
2011, for a total lifetime of 696.32 days.  No evidence could be found for
neutrino bursts from gravitational stellar collapses over the whole period
under study. Considering the null results from previous measurements, it is
concluded that no neutrino burst candidate has been found over 6314 days of
lifetime, during which LVD has been able to monitor the whole Galaxy. A 90\%
c.l.\ upper limit on the rate of gravitational stellar collapses in the Galaxy
($D\le20$~kpc) has been established at 0.13 events/year.

\Section{HE 2.3 Neutrino telescopes and neutrino astronomy}
\label{he23}

Large-scale neutrino telescopes, such as IceCube or ANTARES, are
\Cerenkov\ detectors deployed deep in the Antarctic ice or in the ocean.

\subsection{Atmospheric neutrino (and muon) flux}
Most of the particles registered with water/ice \Cerenkov\ telescopes are muons
and neutrinos from air showers, induced by high-energy cosmic rays in the
atmosphere. The basic idea is that downgoing muons are interpreted as muons
from air showers above the detector, and muons moving upwards are interpreted
as products of neutrinos generated in air showers on the other side of the
Earth.

Relativistic muons passing through matter lose energy by various processes. The
most common one is ionization of water molecules. Above 1~TeV, radiative
processes start to dominate, leading to an almost linear dependence of the
energy loss per unit length on the energy of the muon. Pair production,
bremsstrahlung, and photo-nuclear interactions are the processes that are
responsible for the increase of energy losses at higher energies.

\begin{figure}[t]
 \includegraphics[width=\columnwidth]{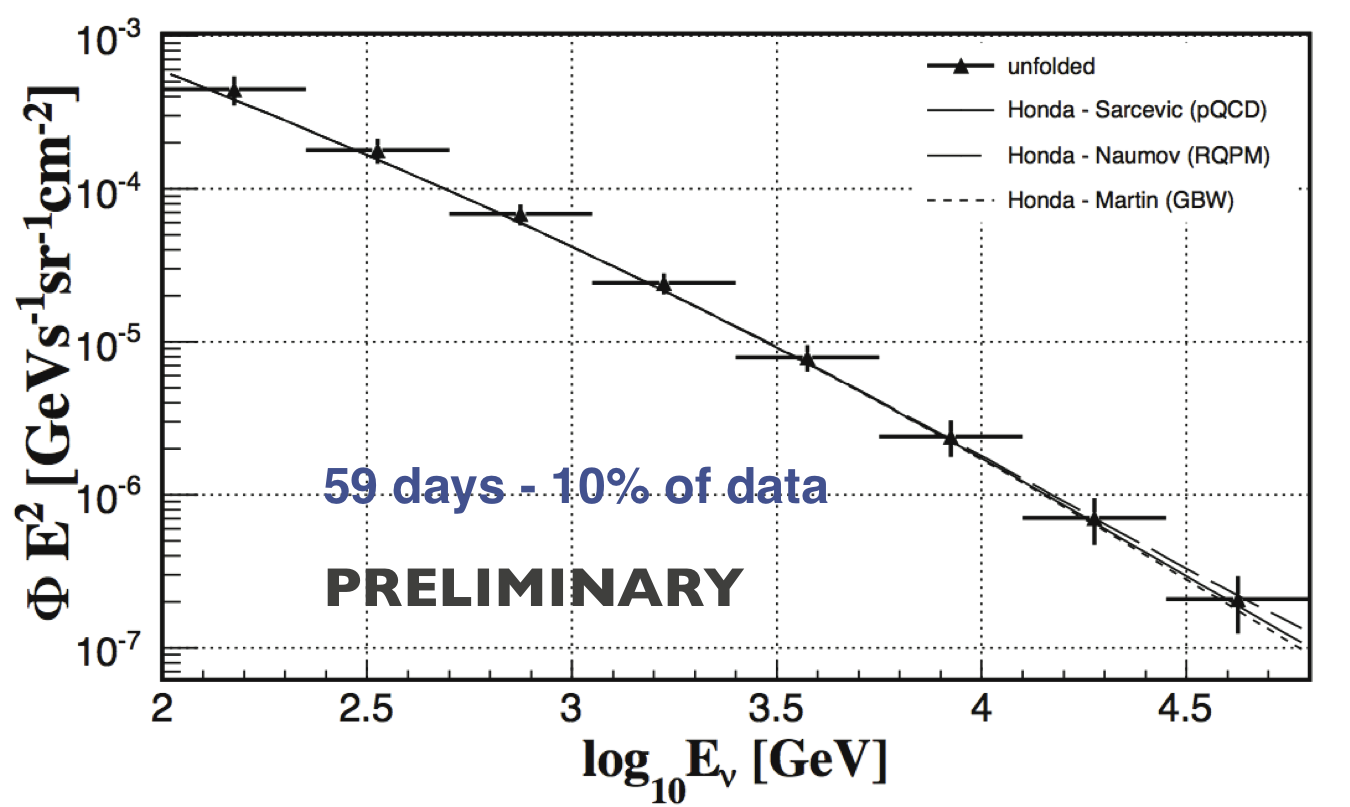}
 \caption{Atmospheric neutrino spectrum as measured by ANTARES
          \iref{Palioselitis}{541}.}
 \label{antares-atmos}
\end{figure} 

The IceCube collaboration has measured the energy spectrum of downgoing
(atmospheric) muons for energies from about 1~TeV up to around 500~TeV
\iref{Berghaus}{85}.  A method has been developed to identify TeV-scale
catastrophic energy losses along muon tracks. It is applied to the separation
of single high-energy muons from large-multiplicity bundles that dominate the
event sample above the horizon at high energies. The information is used to
derive the single-muon energy spectrum at all zenith angles.  The observed muon
flux is sensitive to the composition of cosmic rays.  The measured energy
spectrum of stochastic losses was compared to simulations based on various
primary composition models \cite{pg}.  The measurements are compatible with a
cutoff of the proton flux at the knee in the energy spectrum of cosmic rays.

The ANTARES collaboration also investigates the sensitivity of the down-going
muon flux to the composition of cosmic rays \iref{Hsu}{679}. Several
observational parameters are combined in a multiple-layered neural network to
estimate the relative contribution of light and heavy cosmic rays.

The energy losses of muons are used by the ANTARES Collaboration to derive the
corresponding muon energy \iref{Palioselitis}{541}. An unfolding method is
applied to derive the spectrum of atmospheric neutrinos from the measured muon
flux. The result is depicted in \fref{antares-atmos}.

Neutrinos in air showers originate from the decay of $\pi$ and $K$ mesons (the
conventional flux) and from the decay of charmed mesons (the prompt flux).  At
TeV energies, where atmospheric neutrinos are an inevitable source of
background events for astrophysical neutrino searches, the prompt flux becomes
important and the flux predictions vary greatly.  IceCube has measured the
conventional atmospheric muon neutrino flux \cite{Abbasi:2010ie}.  In addition,
methods are developed to search for prompt neutrinos \iref{Middell}{1097}. Data
analysis with 40 strings of IceCube is under way.

\begin{figure}[t]
 \includegraphics[width=\columnwidth]{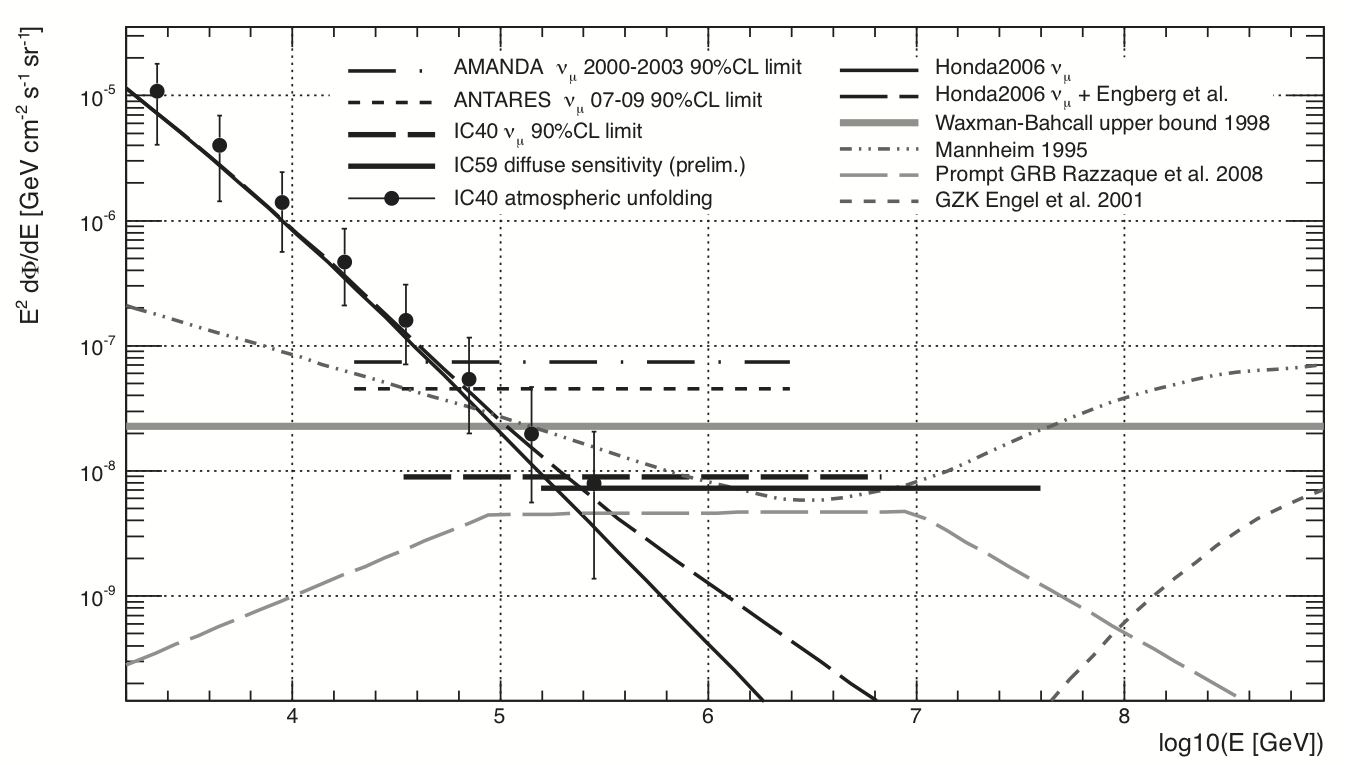}
 \caption{Limit on the diffuse neutrino flux derived by IceCube
          \iref{Schukraft}{736}.}
 \label{icecube-diffuse}
\end{figure} 

\subsection{Diffuse neutrino flux}
The search for a cumulative flux of high-energy neutrinos from the sum of all
cosmic sources in the Universe is one of the central goals of neutrino
telescopes. The experimental signature of isotropically distributed
astrophysical sources is an excess of high-energy neutrinos with a
characteristic angular distribution over the background of less energetic
neutrinos produced when cosmic rays interact with the Earth's atmosphere.  Such
searches are challenging because of systematic uncertainties in these fluxes
and the detector response. 

The measured flux of high-energy cosmic rays has been used to derive upper
bounds on the expected diffuse neutrino flux
\cite{Waxman:1998yy,Bahcall:1999yr,Mannheim:1998wp}. In the TeV to PeV energy
range, this flux is typically assumed to originate from
particle interactions at or close to the cosmic-ray acceleration sites.
Although only weakly constrained, the neutrino energy spectrum is typically
modeled by a simple power law $\propto E^{-2}$. Therefore, limits on the
diffuse neutrino flux are typically presented by multiplying the flux with
$E^2$.

The distribution of reconstructed neutrino energies in IceCube is analyzed
using a likelihood approach that takes into account these uncertainties and
simultaneously determines the contribution of an additional diffuse
extraterrestrial neutrino component \iref{Schukraft}{736}. This analysis is
applied to the data measured with the IceCube detector in its 40- and 59-string
configurations, covering the period from April 2008 to May 2010. No evidence
for an astrophysical neutrino flux was found in the 40-string analysis. The
upper limit obtained for the period from April 2008 to May 2009 is 
$ d\Phi/dE \le 8.9\cdot 10^{-9} ~\mbox{GeV}^{-1} 
     ~\mbox{cm}^{-2}~\mbox{s}^{-1}~\mbox{sr}^{-1}$ 
at 90\% c.l.\ in the energy region between 35~TeV and 7~PeV. For the
59-string data from May 2009 to May 2010, an improved analysis technique has
been developed. The preliminary sensitivity is 
$d\Phi/dE \le 7.2\cdot10^{-9}~\mbox{GeV}^{-1} ~\mbox{cm}^{-2} ~\mbox{s}^{-1}
     ~\mbox{sr}^{-1}.$
The flux limits are shown in \fref{icecube-diffuse}.

\begin{figure}[t]
 \includegraphics[width=\columnwidth]{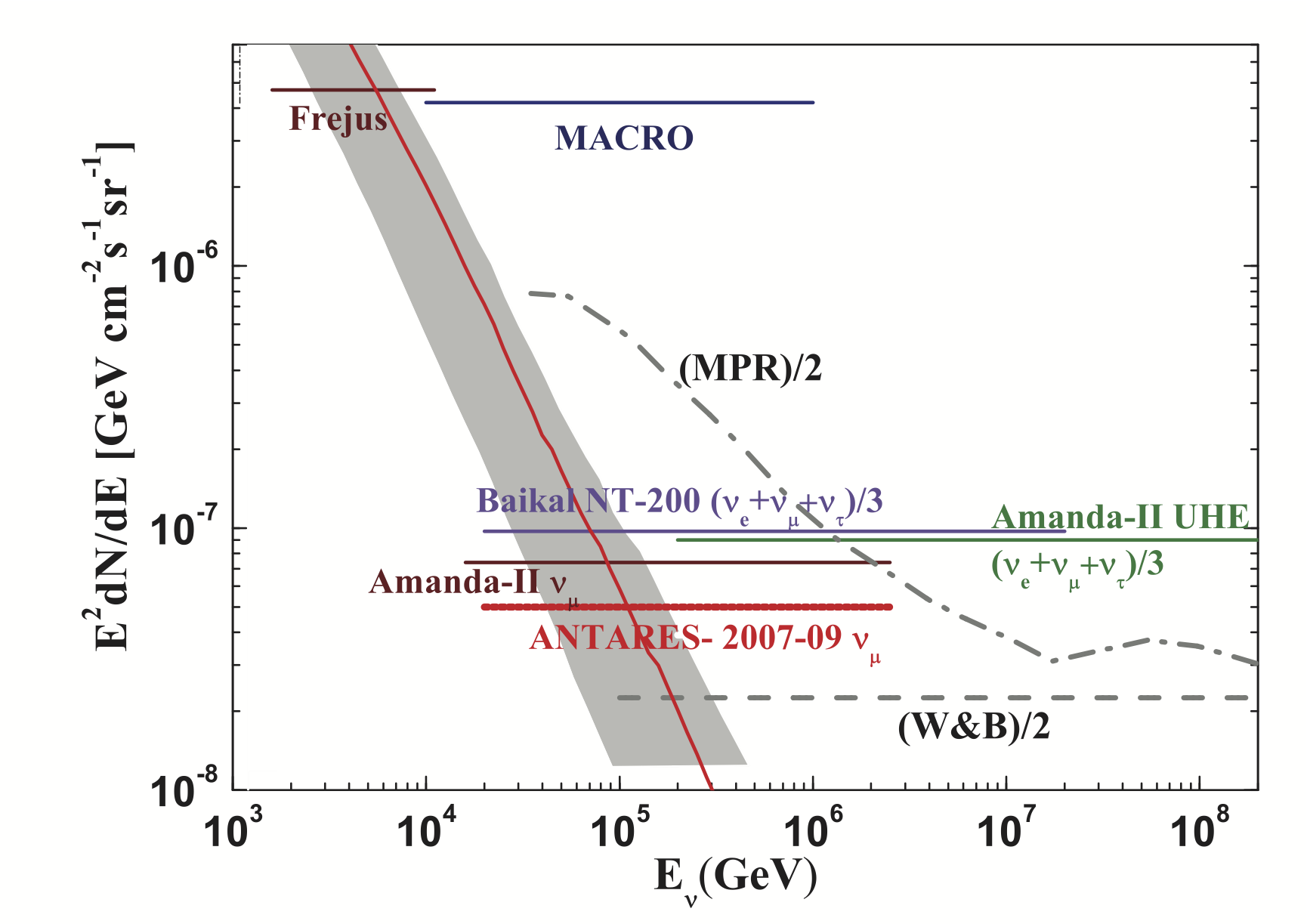}
 \caption{The upper limit for a $E^{-2}$ diffuse high-energy $\nu_\mu$ and
     $\bar{\nu}_\mu$ flux derived from data of the ANTARES neutrino telescope
     compared to previous measurements and phenomenological upper bound
     predictions \iref{Sch\"ussler}{237}.}
 \label{antares-diffuse}
\end{figure} 

Recent results on upper limits for the diffuse neutrino flux obtained by
ANTARES \iref{Sch\"ussler}{237} \iref{Heijboer}{858} are shown in
\fref{antares-diffuse}.  ANTARES obtained an upper limit of 
$ E^2 \Phi_{90\%} = 5.3\cdot 10^{-8}~\mbox{GeV}~
     \mbox{cm}^{-2}~\mbox{s}^{-1}~\mbox{sr}^{-1} . $
The recent upper limits are in the range where
neutrinos are expected according to the above mentioned calculations in the
energy range $\approx 0.1-10$~PeV.

The Surface Detector of the Pierre Auger Observatory can also be used to detect
ultra-high-energy neutrinos in the sub-EeV energy range and above
\iref{Guardincerri}{682}.  Neutrinos of all flavors can interact in the
atmosphere and induce inclined showers close to the ground (down-going). The
sensitivity of the Surface Detector to tau neutrinos is further enhanced
through the "Earth-skimming" mechanism (up-going). Both types of neutrino
interactions can be identified through the broad time structure of the signals
induced in the Surface Detector stations.  Two independent sets of
identification criteria were designed to search for down and up-going neutrinos
in the data collected from 1 January 2004 to 31 May 2010, with no candidates
found. Assuming a differential flux $\Phi(E_\nu) = k E^{-2}$, a 90\% c.l.\
upper limit on the single-flavor neutrino flux has been derived of $k < 2.8
\cdot 10^{-8}$ GeV cm$^{-2}$ s$^{-1}$ sr$^{-1}$ in the energy interval $\approx
10^{17} - 10^{19}$~eV based on Earth-skimming neutrinos and $k < 1.7 \cdot
10^{-7}$ GeV cm$^{-2}$ s$^{-1}$ sr$^{-1}$ in the energy interval $\approx
10^{17} - 10^{20}$~eV based on down-going neutrinos. 

\subsection{Point sources}
The observation of a deficit of cosmic rays from the direction of the Moon is
an important experimental verification of the absolute pointing accuracy of a
neutrino telescope and the angular resolution of the reconstruction methods.
Both, ANTARES \iref{Riviere}{98} and IceCube \iref{Boersma}{1235} have reported
an observation of the shadow of the Moon in the flux of down-going muons.

Several methods have been presented to search for (astrophysical) point sources
in the measured arrival directions of neutrinos.

Clustering of neutrino arrival directions would provide hints for their
astrophysical origin. The two-point autocorrelation method is sensitive to a
large variety of cluster morphologies and provides complementary information to
searches for the astrophysical sources of high-energy muon neutrinos.  The
autocorrelation function has been investigated as a function of the angular
scale of data collected during $2007-08$ with the ANTARES neutrino telescope
\iref{Sch\"ussler}{238}. The data, taken during the deployment phase of the
detector, do not show evidence for deviations from the isotropic arrival
direction distribution expected for the background of atmospheric neutrinos and
contamination by mis-reconstructed atmospheric muons.

\begin{figure}[t]
 \includegraphics[width=\columnwidth]{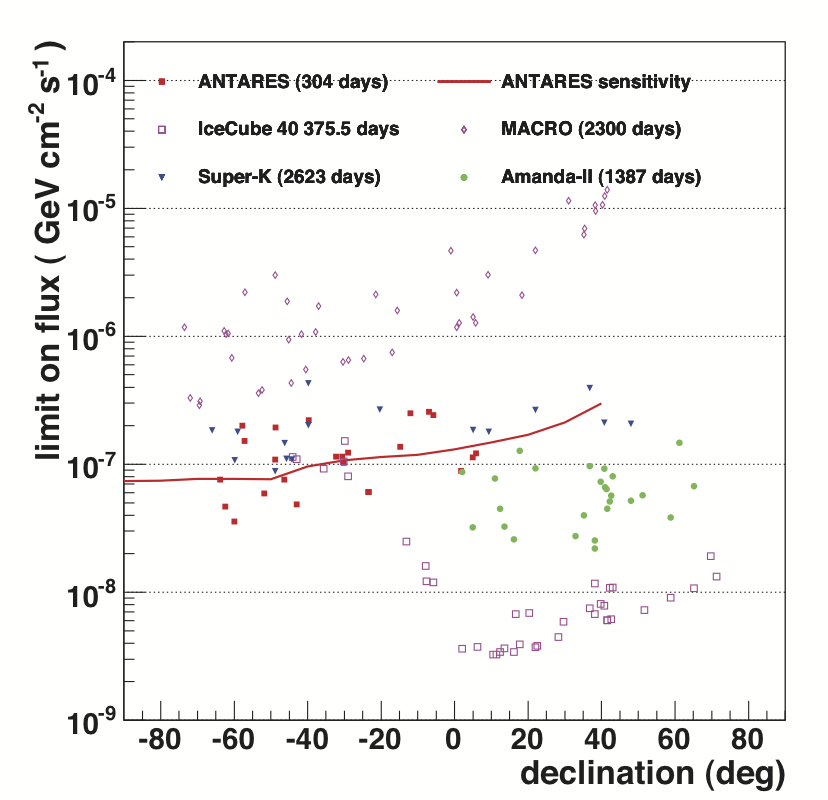}
 \caption{Limits set on the normalization of an $E^{-2}$ spectrum of
	  high-energy neutrinos from selected source candidates as obtained by
          ANTARES \iref{Bogazzi}{295}.}
 \label{antares-point}
\end{figure} 

The first two years of ANTARES data ($2007-08$) have been used to search for
point sources \iref{Bogazzi}{295} \iref{Gomez-Gonzalez}{678}.  No statistically
significant excess of events has been found either in the search using a
candidate list of 24 interesting (gamma-ray) sources, or in a full-sky search.
The obtained flux limits are plotted as function of declination in
\fref{antares-point}.  The most significant cluster, with a post-trial
probability of 18\%, corresponds to the location of HESS J1023-575. 

The data from IceCube-40 and AMANDA, taken during $2008-09$, are used to search
for neutrino sources within the Galaxy  \iref{Odrowski}{320}.
The TeV gamma-ray spectra of some potential Galactic cosmic-ray
accelerators show cut-offs in the energy spectrum at energies of a few TeV. In
the case of transparent TeV gamma-ray sources, high-energy neutrinos will have
similar spectra, and an improved effective area below a few TeV improves the
sensitivity for these sources.
Several tests, including a scan of the Galactic plane in the Northern Hemisphere
and a dedicated analysis for the Cygnus region, have been performed. 
In the absence of a significant signal, upper limits have been obtained.\\
The strongest preliminary flux limit can be set for Cas A at a flux of
$5.9\cdot10^{-11}$~TeV$^{-1}$ cm$^{-2}$ s$^{-1}$.\\
With 55 events observed within the box defined around the most active part of
the Cygnus region compared to a background expectation of 60 events, strong
flux upper limits could be extracted for this region. Assuming an $E^{-2.6}$
spectrum, as obtained from the MILAGRO gamma-ray observations, a preliminary
90\% flux upper limit of $3\cdot10^{-11}$~TeV$^{-1}$ cm$^{-2}$ s$^{-1}$
(without systematic uncertainties) is obtained, provided the astrophysical
signal from the region has an exponential energy cutoff at or above 1 TeV.\\
The results provide the most restrictive upper limits for the Cygnus region
obtained so far. Depending on the assumed energy cut-off, the upper limits
obtained with this analysis are only a factor of two to three above the
expected neutrino flux if all the TeV gamma-rays observed in the region were of
hadronic origin. This implies that during the coming years, IceCube will be
able to either detect neutrinos from the Cygnus region, or to constrain the
nature of the high-energy gamma-ray emission in the region and, thus, the
fraction of interacting cosmic rays produced in one of the most active parts of
the Galaxy.

\begin{figure}[t]
 \includegraphics[width=\columnwidth]{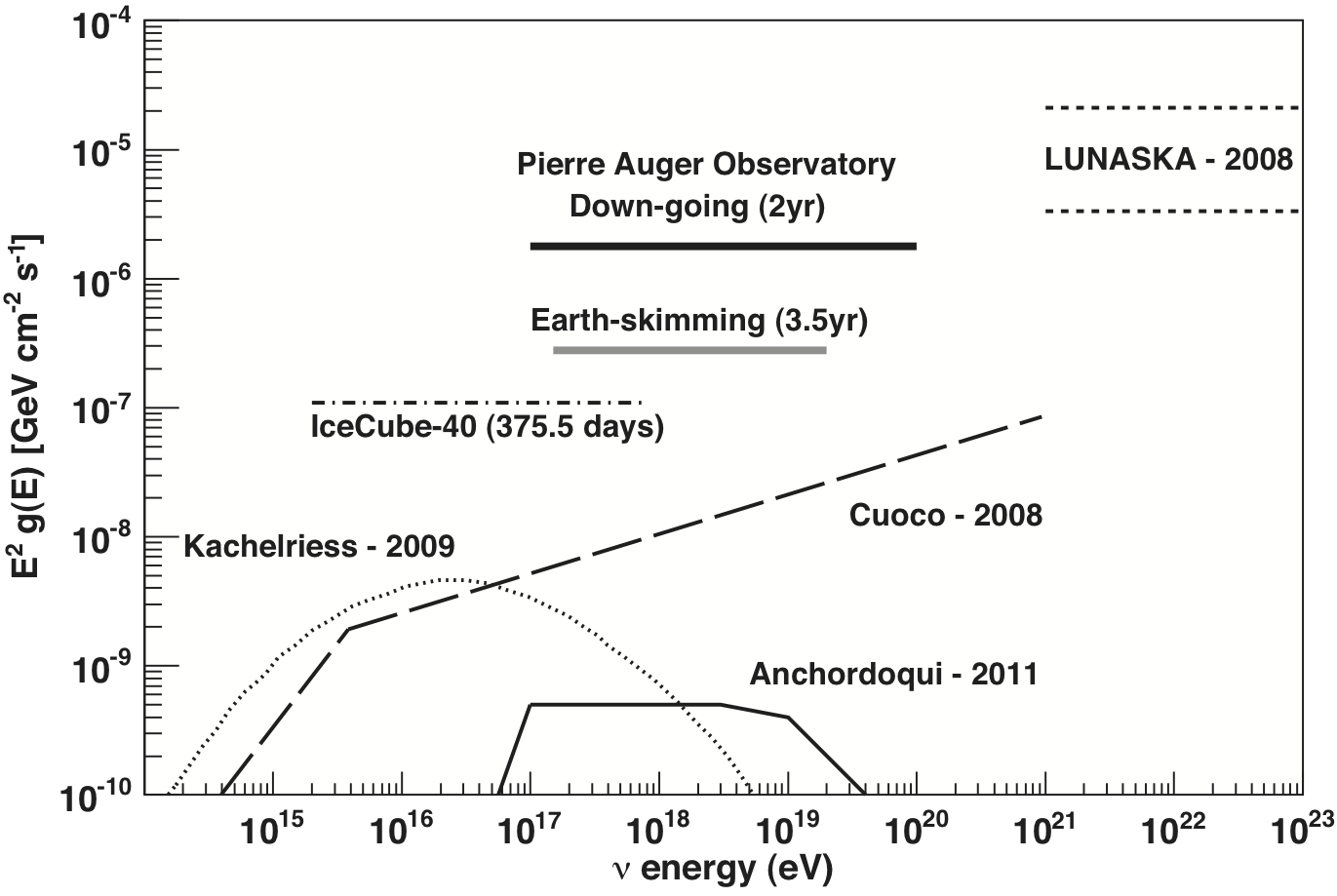}
 \caption{Single-flavor neutrino flux limits (90\% c.l.) on Centaurus A as
          obtained by the Pierre Auger Observatory compared to model 
          predictions\iref{Guardincerri}{682}.}
 \label{auger-cena}
\end{figure} 

The Pierre Auger Observatory has also placed limits on the neutrino flux from
point-like sources as a function of declination \iref{Guardincerri}{682}.  In
particular, an upper limit has been set on neutrinos from the active galaxy
Centaurus A. This Active Galactic Nucleus (AGN) is an interesting source, since
there seems to be an enhancement of charged cosmic rays from the direction of
this source \cite{:2010zzj}.  The obtained 90\% c.l.\ upper neutrino flux limit
is shown in \fref{auger-cena}.  The present upper limits are still about 2
orders of magnitude above the expectations from theoretical models for this
source.

\subsection{Multi messenger searches}
The correlation of information from different messenger particle types is of
great astrophysical interest to identify the sources of high-energy particles.
According to the standard theory of acceleration in astrophysical sources,
charged (hadronic) particles are accelerated in the source and are expected to
be accompanied by gamma-rays and neutrinos from pion decays formed in the
interactions of hadrons with the surrounding medium at the sources.  While
gamma-rays have been linked to astrophysical sources by many experiments
(H.E.S.S, MAGIC, VERITAS, Fermi), no point source of high-energy neutrinos has
been found so far.  Several investigations have been conducted, combining the
search for high-energy neutrinos with photons, charged cosmic rays, and
gravitational waves.

Neutrino multiplets observed with IceCube are used to send a trigger to the
Robotic Optical Transient Search Experiment, ROTSE  \iref{Franckowiak}{445}.
The four ROTSE telescopes immediately observe the corresponding region in the
sky in order to search for an optical counterpart to the neutrino events. Data
from the first year of operation of the optical follow-up program have been
searched for a signal from supernovae. No statistically significant excess in
the rate of neutrino multiplets has been observed, and furthermore, no
coincidence with an optical counterpart was found during the first year of
data-taking. This restricts current models, that predict a high-energy neutrino
flux from soft jets in core-collapse supernovae.

ANTARES is used to initiate optical follow-up observations to search for
transient sources, such as gamma ray bursters (GRBs) and core-collapse
supernovae \iref{Ageron}{90}.  A fast online muon track reconstruction is used
to trigger a network of small automatic optical telescopes. Such alerts are
generated one or two times per month for special events, such as two or more
neutrinos coincident in time and direction or single neutrinos of very high
energy. Since February 2009, ANTARES has sent 37 alert triggers to the TAROT
and ROTSE telescope networks and 27 of them have resulted in followup
observations. So far, no optical counterpart has been detected.

IceCube has also been used to trigger the SWIFT x-ray satellite
\iref{Homeier}{535}.  SWIFT then scans the sky in the direction of the neutrino
for a transient x-ray counterpart, e.g.\ an x-ray gamma-ray burst afterglow.
The program started in February 2011. Until May 2011, one trigger has been
forwarded to SWIFT. The total latency between the neutrino events and the first
observation by SWIFT was 90 minutes.

Finally, there are also plans to use IceCube as trigger for imaging atmospheric
\Cerenkov\ telescopes \iref{Franke}{334}. A test run has been conducted in
early 2011. Interesting events have been identified, but triggers are (not yet)
forwarded to a ground-based \Cerenkov\ telescope.

In the reverse direction, data from the GRB coordinates network
\cite{grbnetwork} were used to initiate a search for neutrinos with IceCube
\iref{Redl}{764}.  The analysis of data from the IceCube 59-string
configuration is a dedicated search for neutrinos produced via $p-\gamma$
interactions in the prompt phase of the GRB fireball. Yielding no significant
excess above the background, this constrains current models of GRBs.

ANTARES data are used to search for neutrinos from known astrophysical
objects.  Radio-loud active Galactic nuclei with their jets pointing almost
directly towards the observer, the so-called blazars, are particularly
attractive potential neutrino point sources \iref{Dornic}{91}.  The gamma-ray
light curves of blazars measured by the Fermi LAT instrument reveal time
variability information.  An unbinned method based on the minimization of a
likelihood ratio was applied to a subsample data collected in 2008 (61 days
lifetime), by looking for neutrinos detected in the high state periods of the
AGN light curve.  The search has been applied to ten very bright and variable
Fermi LAT blazars.  The most significant observation of a flare is 3C279, with
a chance probability of about 10 \% after trials, for which one neutrino event
has been detected in time/space coincidence with the gamma-ray emission. Limits
have been obtained on the neutrino fluence for the ten selected sources. 

In 2007 ANTARES consisted of five detector lines.  These data have been used to
search for coincidences between the observed neutrinos and 37 GRBs
\iref{Reed}{1085}. No correlations have been observed.

\begin{figure}[t]
 \includegraphics[width=\columnwidth]{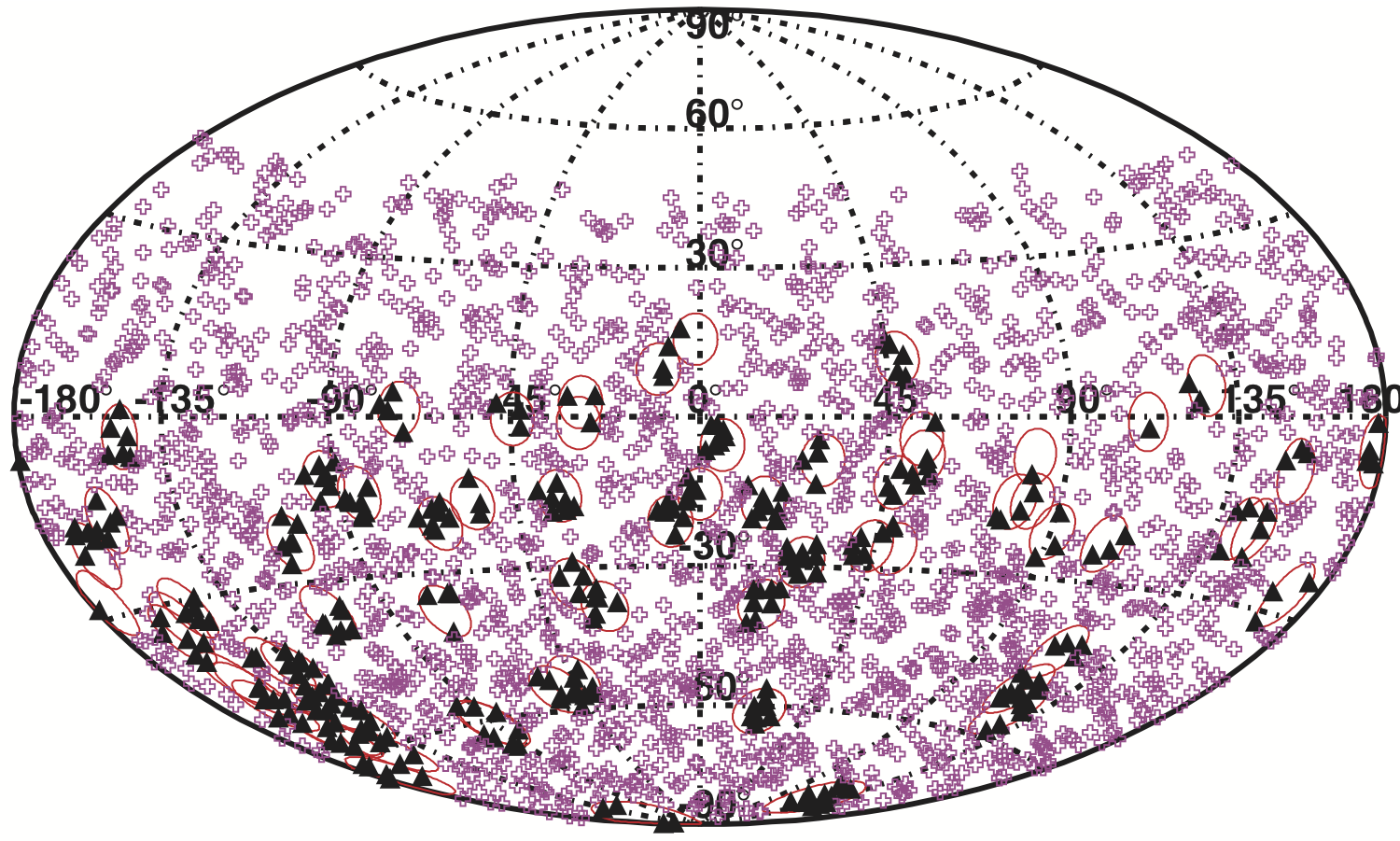}
 \caption{Skymap showing the arrival direction of neutrinos observed by ANTARES
	  (crosses) and cosmic rays as measured by the Pierre Auger
          Observatory (triangles) \iref{Petrovic}{291}.}
 \label{petrovic}
\end{figure} 
A search for correlations between high-energy neutrinos and (charged) cosmic
rays has been performed \iref{Petrovic}{291}.  A source-stacking analysis has
been developed and applied to neutrino-candidate events detected during
$2007-08$ with the ANTARES neutrino telescope and 69 ultra-high-energy cosmic
rays observed by the Pierre Auger Observatory. The corresponding arrival
directions are depicted in a skymap in \fref{petrovic}.  The observed number of
correlated neutrino events is below expectations (negative correlation), with a
significance of about $1.8\sigma$. This result is compatible with a background
fluctuation. The corresponding upper flux limit, assuming an equal flux from
all ultra-high-energy cosmic-ray sources, is
$5.43\cdot10^{-8}$~GeV~cm$^{-2}$~s$^{-1}$.

Cataclysmic cosmic events can be plausible sources of both, gravitational waves
and high-energy neutrinos. Both are alternative cosmic messengers that may
traverse very dense media and travel unaffected over cosmological distances,
carrying information from the innermost regions of the astrophysical engines.
Such messengers could also reveal new, hidden sources that were not observed by
conventional photon astronomy \iref{van Elewyck}{701}.
A joint search was conducted with the ANTARES neutrino telescope and the
LIGO/VIRGO gravitational wave detectors using concomitant data taken in 2007
during the VIRGO VSR1 and LIGO L5 science runs, while ANTARES was operating in
a 5-line configuration. No coincident events have been found so far.

\Section{HE 2.4 Theory and calculations}\label{he24}

\subsection{The endpoint formalism}
\begin{figure}[t]
 \includegraphics[width=\columnwidth]{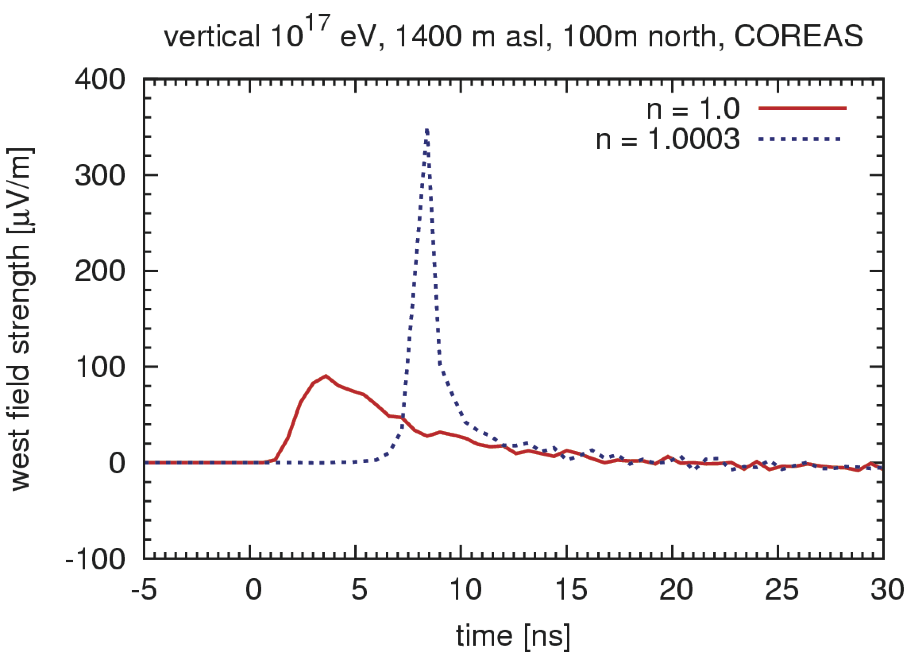}
 \caption{Radio emission from air showers.
          Expected field strength as function of time for media with two
          different refractory indices \iref{Huege}{653}.}
 \label{coreas}
\end{figure} 

Huege \etal\ propose a formalism suitable for the calculation of
electromagnetic radiation from any kind of particle acceleration, which lends
itself very well to the implementation in Monte Carlo codes, and necessitates
no simplifying approximations to the underlying processes \iref{Huege}{653}. In
this "end-point formalism", presented in detail in \rref{James:2010vm}, the
trajectory of individual particles is described as a series of points at which
particles are accelerated instantaneously, leading to discrete radiation
contributions which can then be easily superimposed.  The endpoint formalism
can be applied in both, the frequency and time domains, and correctly
reproduces several electromagnetic processes, like synchrotron radiation,
\Cerenkov\ radiation, and transition radiation. Furthermore, the authors
demonstrate how the application of the endpoint formalism provides insights to
radio emission processes in astroparticle physics.

As an example for an application of the method, predictions for the radio
emission in air showers are shown in \fref{coreas}. It depicts the expected
field strength at ground level as function of time for two assumptions for the
refractive index of air. A more realistic index of $n>1$ yields a sharper radio
pulse.

\subsection{Atmospheric neutrino spectrum}
High-energy neutrinos, arising from decays of mesons produced through the
collisions of cosmic-ray particles with air nuclei, form a background for the
detection of astrophysical neutrinos \iref{Sinegovsky}{487}.  An ambiguity in
the high-energy behavior of pion and especially kaon production cross sections
for nucleon-nucleus collisions may affect the calculated neutrino flux.  The
authors present results of the calculation of the energy spectrum and
zenith-angle distribution of muon neutrinos and electron neutrinos of
atmospheric origin in the energy range from 10~GeV to 10~PeV. The calculations
were performed using several hadronic interaction models (QGSJET-II, SIBYLL
2.1, Kimel \& Mokhov) and two parameterizations of the primary cosmic-ray
spectrum, by Gaisser \& Honda and by Zatsepin \& Sokolskaya.  The results
demonstrate a weak dependence on the two models used for the primary cosmic-ray
composition.  The predicted fluxes are compared to measurements by the Frejus,
AMANDA, and IceCube experiments. This comparison indicates that QGSJET-II is
the preferred model to describe the data.  An analytic description of the
calculated neutrino fluxes is given, for details see \iref{Sinegovsky}{487}.

\subsection{Neutrino cross section}
\begin{figure}[t]
 \includegraphics[width=\columnwidth]{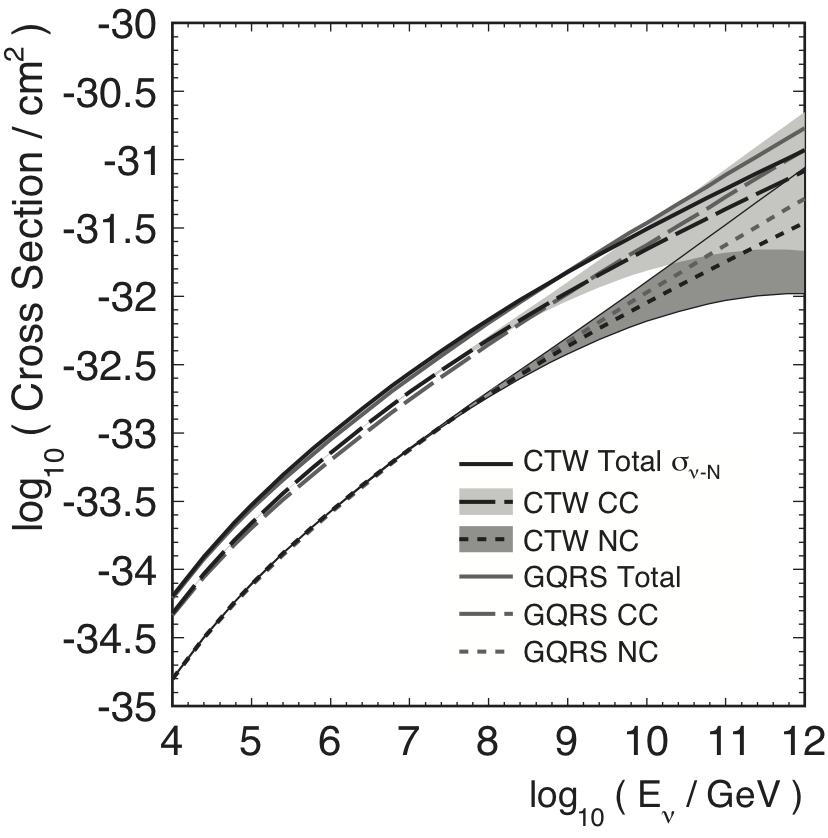}
 \caption{Calculated neutrino-nucleon cross section for charged-current
	  and neutral-current reactions. For details, see
          \cite{Connolly:2011vc} \iref{Connolly}{1283}.}
 \label{nu-crossect}
\end{figure} 
A new calculation has been presented of the cross sections for charged current
and neutral current neutrino-nucleon and anti-neutrino-nucleon interactions in
the neutrino energy range $10^4<E_\nu<10^{12}$~GeV \cite{Connolly:2011vc}
\iref{Connolly}{1283}. The parton distribution functions have been used as
calculated by A.D.\ Martin \etal\ known as "MSTW 2008". The latter incorporates
improvements in the precision and kinematic range of recent measurements as
well as improved theoretical developments that make the global analysis more
reliable.  The calculated neutrino-nucleon cross section is depicted in
\fref{nu-crossect}.  Good agreement is found between the new calculations and
previous results.

\Section{HE 2.5 Muon and neutrino tomography}\label{he25}

\subsection{Geophysical application}
The capability of high-energy atmospheric muons to penetrate large depths of
material makes them a unique probe for geophysical studies. Provided the
topography is known, the measurement of the attenuation of the muon flux
permits the cartography of density distributions, revealing spatial and
possibly also temporal variations in extended geological structures. A
collaboration between volcanologists, astroparticle- and particle physicists
has been formed to study tomographic muon imaging of volcanos with
high-resolution tracking detectors \iref{Fehr}{671}.

To achieve high resolution and low noise conditions, a muon telescope has been
constructed, consisting of three planes of glass resistive plate chambers. A
1~m$^2$ chamber is read out by 9142 electronic channels.

\begin{figure}[t]
 \includegraphics[width=\columnwidth]{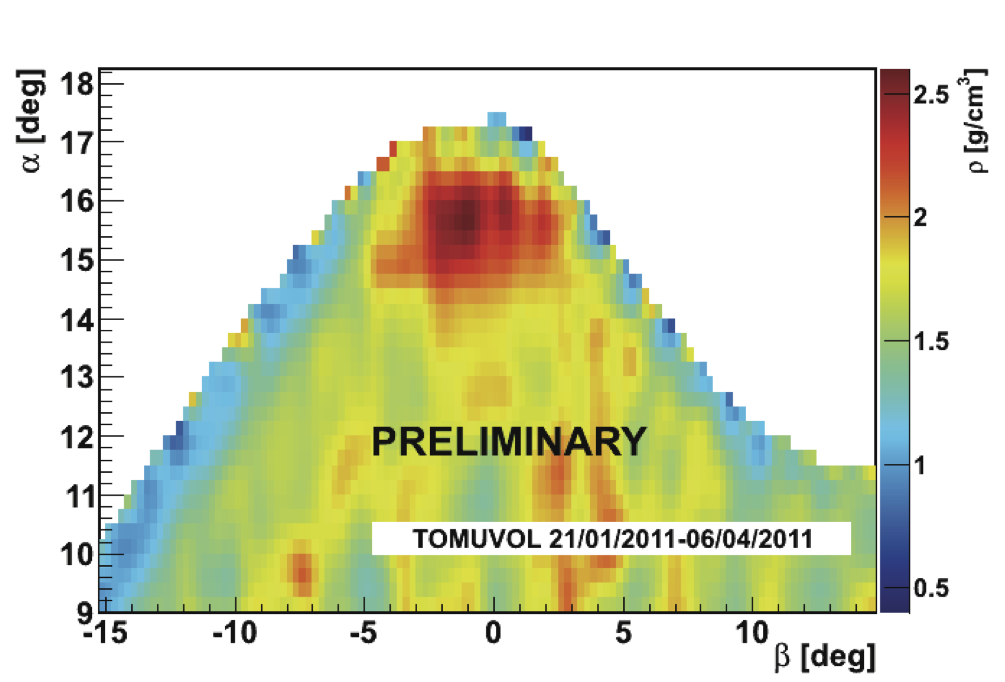}
 \caption{Measured average matter densities within the Puy de Dome along
	  trajectories crossing the detection site 2~km off and 600~m below
          the summit. The measurement is based on 65.8 days of data-taking 
          \iref{Fehr}{671}.}
 \label{vulcano}
\end{figure} 
First measurements have been conducted at the Puy de Dome, an inactive volcanic
dome in south-central France. The shadow cast in the atmospheric muon flux has
been observed. A reference region in the open sky is used to define an
unabsorbed muon flux. This reference flux can be used to determine the
absorption of muons in the volcano. The column density along trajectories
crossing the detection site can be extracted, see \fref{vulcano}.  The density
map exhibits two features: a band of lower density at the flanks and a region
with higher density just beneath the summit.

\subsection{Archaeological application}
The Pyramid of the Sun, at Teotihuacan, Mexico, is being searched for possible
hidden chambers, by means of muon attenuation measurements inside the pyramid's
volume \iref{Menchaca-Rocha}{1117}.  A muon tracker is located in a tunnel,
running below the base and ending close to the symmetry axis of the monument.
The experimental set-up, with an active area of 1~m $\times$ 1~m consists of
six layers of 200-wire multiwire proportional chambers, interspaced with four
layers of plastic scintillator planes, read out by wavelength shifter bars.

The muon transport through the pyramid is simulated using GEANT4. The pyramid
shape is constructed from aerial photographs.  The muon flux inside the pyramid
has been measured for six months and compared to expectations from the
simulations. The comparisons show qualitative and quantitative resemblances.
Given the limited statistics accumulated so far, it is still too early to
confirm or disprove the possibility of a human-size hidden empty cavity within
the pyramid's volume.

\Section{HE 2.6 New experiments and instrumentation}\label{he26}
The status of several endeavors to build new detectors for the measurement of
high-energy neutrinos have been presented.

Dedicated high-energy neutrino telescopes based on optical \Cerenkov\
techniques have been scanning the cosmos for about a decade. Consequently,
neutrino flux limits have improved by several orders of magnitude in the
TeV-PeV energy interval. At higher energies, detectors using radio \Cerenkov\
techniques have produced aggressive limits on the neutrino flux. 

\subsection{KM3NeT} 
\begin{figure}[t]
 \includegraphics[width=\columnwidth]{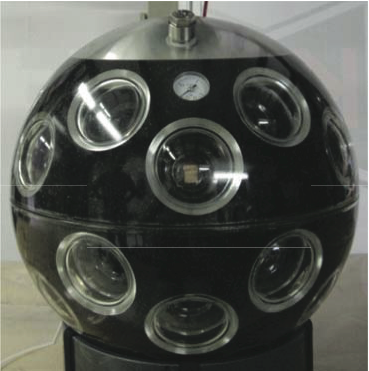}
 \caption{Digital Optical Module (DOM) of the KM3NeT neutrino telescope. It
    consists of photomultipliers with light concentrator rings, 
    foam support, an aluminum cooling structure, and a  cable
    feed-through \iref{Kooijman}{800}.}
 \label{km3net-dom}
\end{figure}
The KM3NeT project is aiming to construct a multi-km$^3$ water \Cerenkov\
neutrino telescope \iref{Kooijman}{800}.  Its envisaged location in the deep
Mediterranean Sea optimizes the sensitivity for neutrino sources near the
Galactic center. In 2010/11 a number of major decisions have been made on the
technical implementation. A design has been adopted based on a digital optical
module (DOM) configuration, housing many small photomultipliers rather than
earlier traditional designs, having a single large PMT.  A DOM is depicted in
\fref{km3net-dom}.  The readout electronics are incorporated into this digital
optical module.  The site-to-shore communication is provided by an
electro-optical cable, allowing unique color point-to-point communication
between each DOM and the shore station readout system.  There are plans to
install a DOM within ANTARES for an in-situ test under realistic environmental
conditions \iref{Popa}{386}.  Each mechanical supporting structure consists of
twenty 6~m long bars with a DOM at each end.  The bars have a vertical
separation of 40~m, giving the full structure a height of around 900~m. The
foreseen budget allows for the construction of about 300 of such structural
units.  Together they will create an instrumented volume of between 5 and 8
km$^3$. The sensitivity to point sources near the Galactic center is expected
to surpass that of earlier telescopes by two orders of magnitude. Construction
is planned to start in 2014.

\subsection{The Askaryan radio array - ARA} 
The Askaryan radio array (ARA) has the objective to detect ultra-high-energy
neutrinos in a dense, radio-frequency transparent medium via the Askaryan
effect \iref{Hoffman}{1316}.  It is built on the expertise gained by RICE,
ANITA, and IceCube's radio extension in the use of the Askaryan effect in cold
Antarctic ice.  The goal is to install an antenna array in bore holes extending
200 m below the surface of the ice near the geographic South Pole. The
unprecedented scale of ARA, which will cover a fiducial area of 200 km$^2$, was
chosen to ensure the detection of the flux of neutrinos expected from the
observation of a GZK-like feature by HiRes and the Pierre Auger Observatory via
cosmic-ray interactions with the cosmic microwave background.

Funding to develop the instrumentation and install the first prototypes has
been granted, and the first components of ARA were installed during the austral
summer of $2010-2011$ \iref{Connolly}{1237}. Within 3 years of commencing
operation, the full ARA will exceed the sensitivity of any other instrument in
the $0.1-10$~EeV energy range by an order of magnitude. The primary goal of ARA
is to establish the absolute cosmogenic neutrino flux through a modest number
of events. 

\begin{figure}[t]
 \includegraphics[width=\columnwidth]{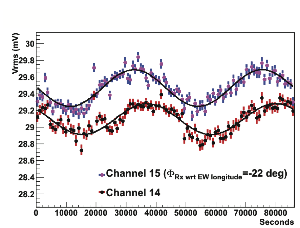}
 \caption{Radio background below 70 MHz as function of time of day as recorded
          by ARA \iref{Conolly}{1237}.}
 \label{ara-gal}
\end{figure}

In January 2011, a first prototype testbed station for ARA was deployed 1.8~km
east of the IceCube detector, at approximately 30 m depth \iref{Conolly}{1237}.
It included ten antennas deployed in the ice and six antennas deployed at the
surface, covering frequencies from 30 to 850 MHz.  The surface antennas show an
increase in thermal noise at frequencies below 150 MHz. This is consistent with
Galactic radio noise, whose sky temperature follows the relation
\begin{equation}
 T_{sky} = 800~\mbox{K}(f/100~\mbox{MHz}) -2.5 ,
\end{equation}
where $f$ is the frequency.  The radio background measured with two antennas is
shown in \fref{ara-gal}.  One recognizes a sidereal modulation of the signal.
Such a behavior is consistent with the observation of radio emission from the
Galaxy, since the two surface antennas are low frequency dipoles with axes
lying parallel to the surface and they sweep through the Galactic plane over a
period of a day. The Galactic plane is at approximately 63$^\circ$ declination
at the South Pole.  The phase difference in the sine waves from the two
antennas is due to an approximately 22$^\circ$ offset in orientation between
them.
The electronics of the currently deployed testbed detector are based on proven
technology from the ANITA balloon experiment and further development of
electronics is in progress \iref{Allison}{744}. 

\subsection{ARIANNA}
A new concept for the detection of GZK neutrinos has been introduced ---
ARIANNA \iref{Barwick}{976}. This next-generation astrophysical neutrino
detector takes advantage of unique geophysical features of the Ross Ice Shelf
in Antarctica. ARIANNA, based on the radio \Cerenkov\ technique, is designed to
improve the sensitivity to neutrinos with energies in excess of $10^{17}$~eV by
at least a factor of 10 relative to current limits. The objective of ARIANNA is
a measurement of the GZK neutrino flux and the search for non-standard particle
physics. 

\begin{figure}[t]
 \includegraphics[width=\columnwidth]{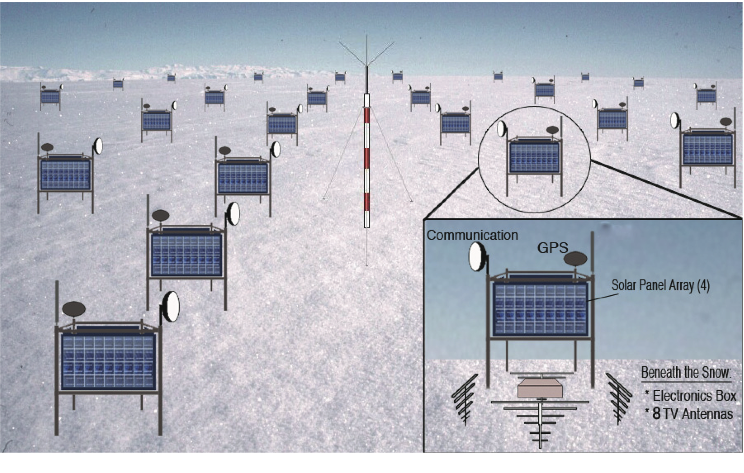}
 \caption{Schematic view of the ARIANNA experiment to measure neutrinos at
          the highest energies \iref{Barwick}{976}.}
 \label{arianna}
\end{figure}

A schematic view of the planned set-up is shown in \fref{arianna}. The
experiment will consist of autonomous radio detection stations on a hexagonal
grid with a separation between stations of the order of 1~km. Each station will
consist of in-ice radio antennas, a GPS receiver for timing, a solar panel
array, and a communication unit.  The first stage of ARIANNA  was approved by
the U.S.\ NSF in mid 2010.

First studies of the site properties have been conducted \iref{Hanson}{340}.
Preliminary results in the frequency band $90-180$~MHz confirm large
attenuation lengths of $(495\pm15$)~m.

\subsection{Acoustic neutrino detection} 
\begin{figure}[t]
 \includegraphics[width=\columnwidth]{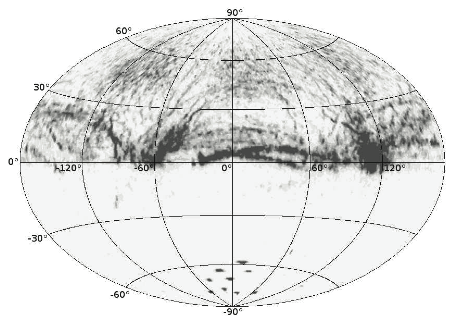}
 \caption{Map of directions of sources reconstructed with an acoustic detector
          (AMADEUS) in the ANTARES neutrino telescope \iref{Lahmann}{894}.}
 \label{amadeus}
\end{figure}
The AMADEUS system is an integral part of the ANTARES neutrino telescope in the
Mediterranean Sea \iref{Lahmann}{894}.  The project aims for the investigation
of techniques for acoustic neutrino detection in the deep sea. Installed at a
depth of more than 2000~m, the acoustic sensors of AMADEUS are based on
piezo-ceramic elements for the broad-band recording of signals with frequencies
up to 125~kHz.  AMADEUS was completed in May 2008 and is composed of six
“acoustic clusters”, each one holding six acoustic sensors that are arranged at
distances of roughly 1~m from each other. The clusters are installed with
spacings ranging from 15~m to 340~m. Acoustic data are continuously acquired
and processed at a computer cluster where online filter algorithms are applied
to select a high-purity sample of neutrino-like signals.  In order to assess
the background of neutrino-like signals in the deep sea, the characteristics of
ambient noise and transient signals have been investigated.

A result of the measurements is depicted in \fref{amadeus}. It shows a map of
directions of reconstructed acoustic sources.  Zero degrees in azimuth
corresponds to the north direction and the polar angle of zero corresponds to
the horizon of an observer on the acoustic storey. At the bottom, the signals
of the emitters of the ANTARES positioning system are visible.  The
measurements of the ambient noise show that the noise level is very stable and
at the expected level, allowing for measurements of neutrino energies down to
$\approx 1$~EeV.

\subsection{Particle interactions on the Moon}
High(est)-energy cosmic rays and neutrinos impinging on the Moon are
expected to emit radio emission in the MHz regime. Such radiation is searched
for with the LOFAR radio telescope \iref{Scholten}{86} \iref{Ter Veen}{1042}.
At energies exceeding $10^{21}$~eV upper limits on the neutrino flux of the
order of $E^2 dN/dE\approx10^{-10}$~GeV~cm$^{-2}$~sr$^{-1}$~s$^{-1}$ are
expected for one week observation time.

\Section{HE 3.1 Hadronic interactions}\label{he31}
The properties of cosmic rays at high energies are investigated through the
measurement of extensive air showers, induced by high-energy particles in the
atmosphere. The understanding of high-energy hadronic interactions is crucial
for a correct interpretation of the air shower data.

\subsection{LHCf}
\begin{figure}[t]
 \includegraphics[width=\columnwidth]{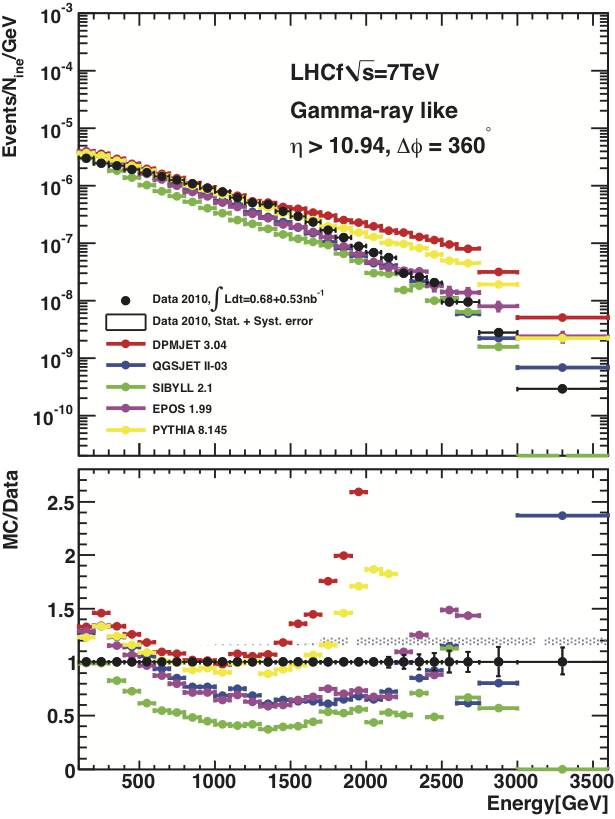}
 \caption{Single-photon energy spectrum in proton-proton interactions as
	  measured by LHCf. The top panels shows the measured values together
	  with predictions from hadronic interaction models. In the lower panel
          the ratio of predictions to measured values are displayed
          \iref{Mitsuka}{1000}.}
 \label{lhcf-photon}
\end{figure}

First results from the Large Hadron Collider (LHC) open insight into a new
energy regime, never covered before at a man-made accelerator.  The energy
range of the LHC is indicated on the upper scale in \fref{ulrich-pair}.  Of
particular interest for cosmic-ray physics is the LHCf experiment. It has two
independent detectors, installed 140~m from the interaction point of the ATLAS
experiment \cite{lhcf}.
The principle idea is to measure charged particles as
close as possible to the beam, investigating the forward direction with small
momentum transfer, i.e.\ large pseudo rapidities $|\eta|>4$.
\footnote{The pseudo rapidity $\eta$ describes the angle of a particle relative
to the beam axis. Is is defined as
$\eta=-\ln\left[\tan\left(\theta/2\right)\right]$, where $\theta$ is the angle
between the particle momentum $\vec{p}$ and the beam axis, or, using the
longitudinal component $p_L$ of the particle momentum
$\eta=1/2\ln\left[(|\vec{p}|+p_L)/(|\vec{p}|-p_L)\right]$.}
The baseline set-up and upgrades of the experiment have been presented
\iref{Suzuki}{264} 
\iref{Noda}{421} 
\iref{Kawade}{959}.
The performance of LHCf was studied with SPS beams in 2007 \iref{Mase}{378}.
A critical parameter for the analysis is the total luminosity, it can be
determined with an accuracy on the $4-5$\% level \iref{Taki}{374}.

\begin{figure}[t]
 \includegraphics[width=\columnwidth]{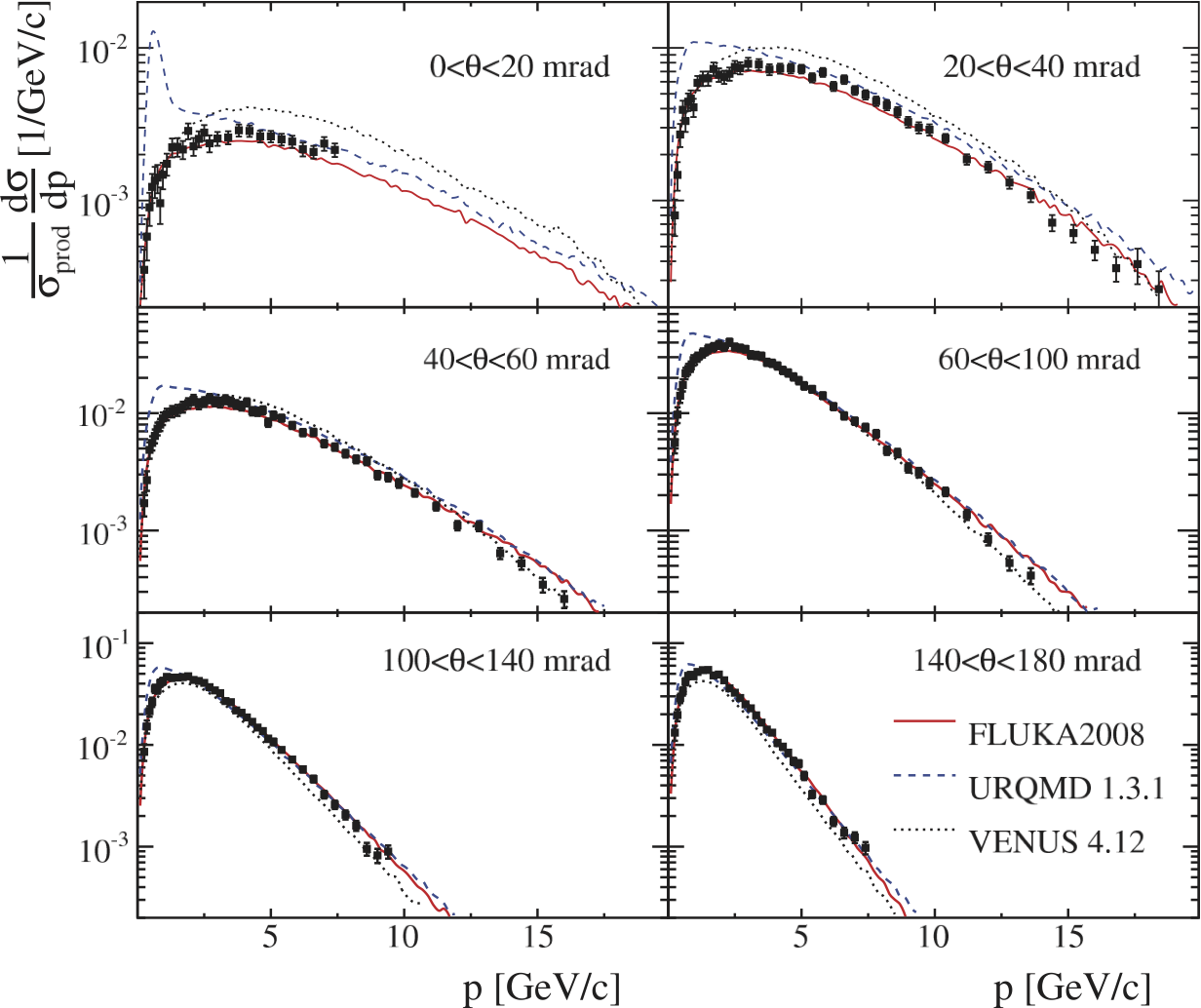}
 \caption{Momentum spectra of $\pi^-$ in p+C interactions at 31~GeV/c measured
	  by NA61 compared to predictions with FLUKA2008, URQMD1.3.1, and
          VENUS4.12 \iref{Unger}{1094}.}
 \label{unger-na61}
\end{figure}

The first phase of data-taking of LHCf has been finished in 2010 with
proton-proton collisions at center-of-mass energies of $\sqrt{s}=0.9$ and 7~TeV
\iref{Sako}{964}.  As an example for recent results, the single-photon energy
spectrum in forward direction ($\eta>10.94$) is presented in \fref{lhcf-photon}
\iref{Mitsuka}{1000}.  The results obtained with the two detectors on either
side of the interaction point are consistent with each other and a combination
of the results of both detectors is shown in the figure.  In the top panel the
measured values are compared to predictions of various hadronic interactions
models used in the simulation of extensive air showers for cosmic-ray physics.
The ratio of model predictions to data is displayed in the lower panel.  It can
be realized that none of the models used describes the measurements perfect.
Such distributions are valuable to tune hadronic interaction models.  Data from
the LHCf experiment are expected to play a crucial role in the further
development of hadronic interaction models in the near future. 

\subsection{NA61/SHINE}
At lower energies (up to 350~GeV) the NA61/SHINE fixed-target experiment is
studying hadron production in hadron-nucleus and nucleus-nucleus collisions at
the CERN SPS \iref{Unger}{1094}. The physics program is optimized for
cosmic-ray physics, e.g.\ investigating collisions of protons and pions with a
carbon target.  As an example for recent results, the momentum spectra of
$\pi^-$ in proton-carbon interactions at a momentum of of 31~GeV/c are shown in
\fref{unger-na61}.  The different panels depict results for different angles
between the secondary pions and the beam axis.  The measured values are
compared to predictions of hadronic interaction models used for air-shower
simulations. Such comparisons are valuable for a further development of
hadronic interaction models.

\subsection{Proton-air and proton-proton cross sections}
\begin{figure}[t]
 \includegraphics[width=\columnwidth]{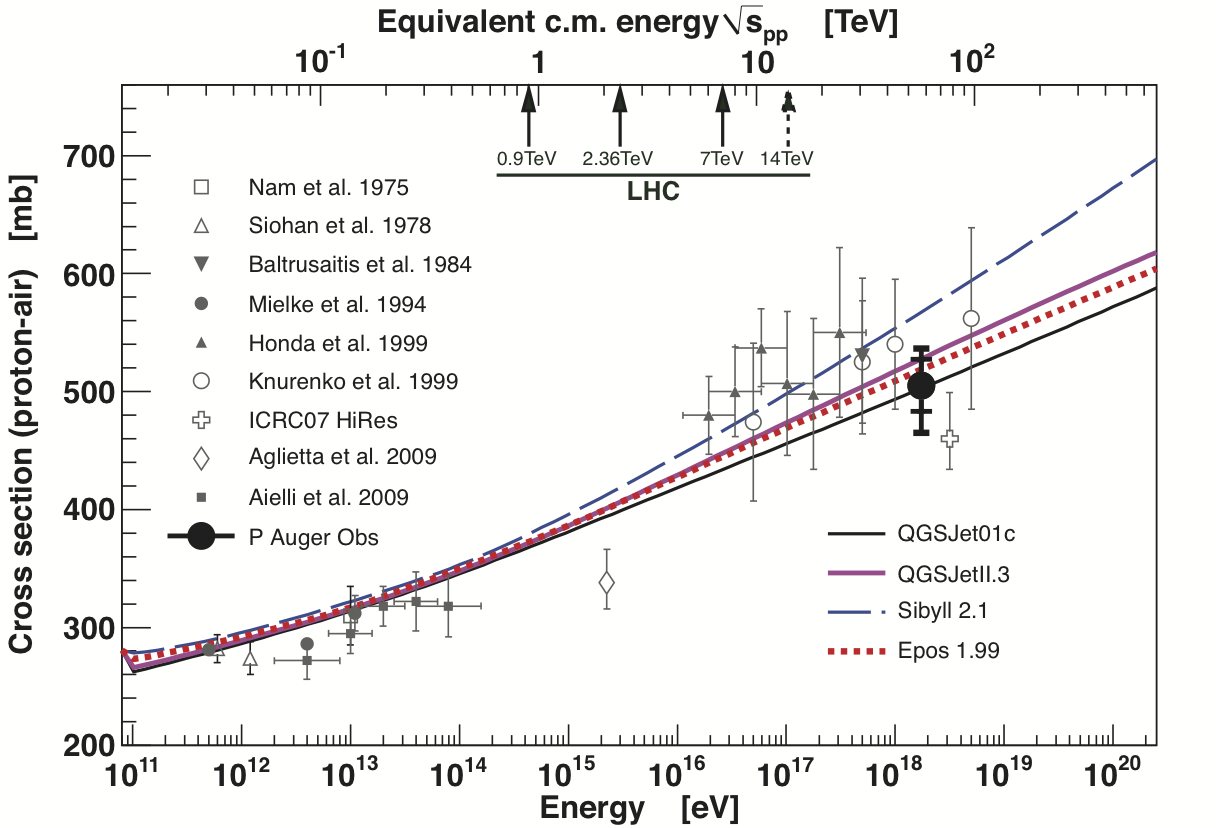}
 \caption{Proton-air cross section derived from air shower measurements with
	  the Pierre Auger Observatory together with results from other
	  experiments and predictions of hadronic interaction models
          \iref{Ulrich}{946}.}
 \label{ulrich-pair}
\end{figure}

Air showers measured with the Pierre Auger Observatory have been used to derive
the proton-air interaction cross section at a
center-of-mass energy of 57~TeV \iref{Ulrich}{946}.  Air showers observed
with the fluorescence detector and at least one station of the Surface Detector
array are analyzed in the energy range from $10^{18}$ to $10^{18.5}$~eV.  A fit
to the tail of the observed distribution of shower maxima yields a value of
$\Lambda=(55.8\pm2.3_{stat}\pm0.6_{syst})$~g/cm$^2$ 
for the attenuation length.  Emphasis is given to systematic uncertainties in
the cross section estimate arising from the limited knowledge of the primary
mass composition, the need to use shower simulations, and the selection of
showers.  For the calculation of the proton-air cross section one of the main
uncertainties is the unknown mass composition of cosmic rays. In particular, a
(unknown) contribution of helium nuclei is expected to bias the obtained
values. This influence has been systematically studied.  For the proton-air
cross section at ($57\pm6$)~TeV a value of
$$ \sigma_{p-air}=\left(505\pm22_{stat} \left(\mbox{\footnotesize +20}
      \atop\mbox{\footnotesize --15} \right)_{syst}\right)~\mbox{mb}$$
has been derived.  The helium-induced systematics is $-12$, $-30$, and
$-80$~mb for a helium contribution of 10\%, 25\%, and 50\%, respectively.  The
bias due to photons in cosmic rays is $<10$~mb.
The value obtained is shown in \fref{ulrich-pair} together with cross sections
obtained in previous experiments.

\begin{figure}[t]
 \includegraphics[width=\columnwidth]{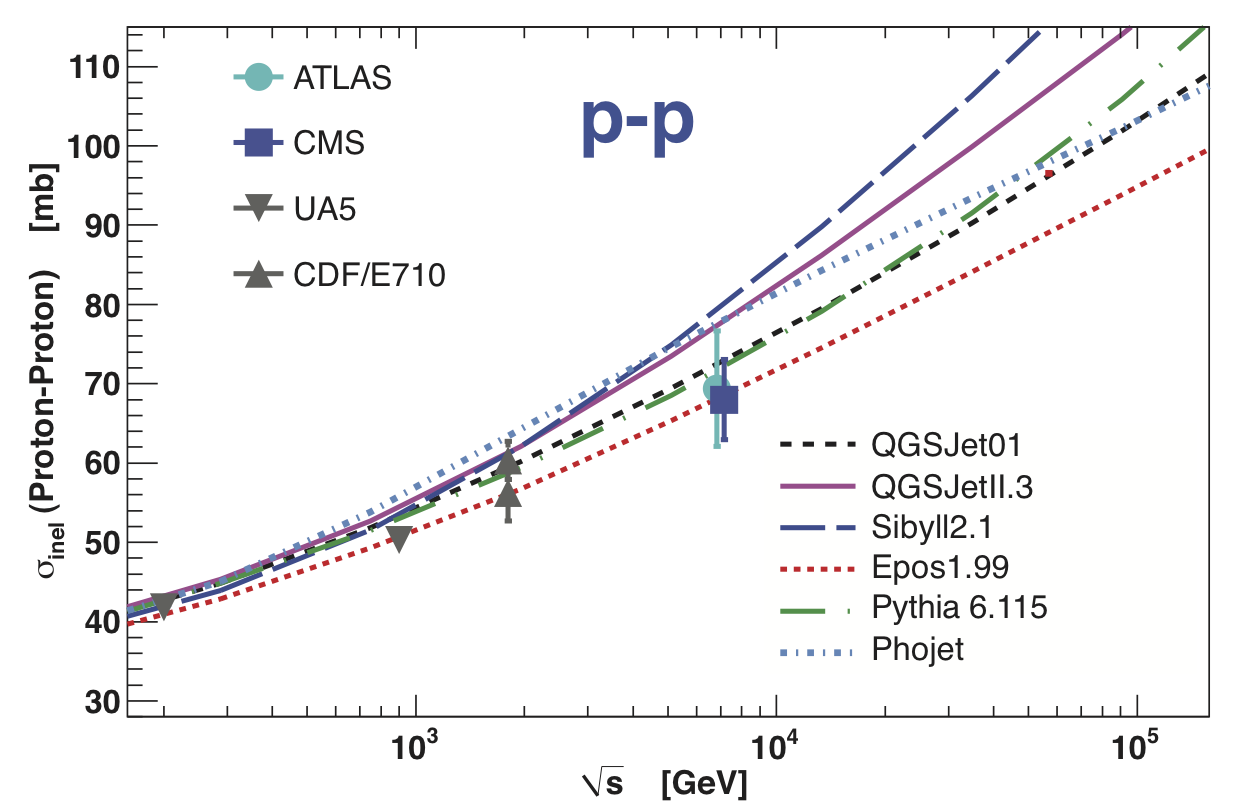}
 \caption{Proton-proton cross section as recently measured at the LHC
	  ($\sqrt{s}=7$~TeV) and earlier results from UA5 and CDF compared to
	  predictions of hadronic interaction models, after
          \iref{Ulrich}{946}.} \label{ulrich-pp}
\end{figure}

The Auger Collaboration is about to derive the proton-proton cross section from
the measured proton-air cross section at $\sqrt{s}=57$~TeV.  Recent
measurements of the proton-proton cross section from the LHC experiments ATLAS
and CMS at a center-of-mass energy of $\sqrt{s}=7$~TeV are shown in in
\fref{ulrich-pp}. 

In Figs.\,\ref{ulrich-pair} and \ref{ulrich-pp} predictions from hadronic
interaction models, used for air-shower simulations are shown in addition to the
experimental values. It should be noted, that the recent Auger results and the
recent LHC measurements are at the lower boundary of the range predicted by the
various interaction models.

\subsection{Air shower data and models}

\begin{figure}[t]
 \includegraphics[width=\columnwidth]{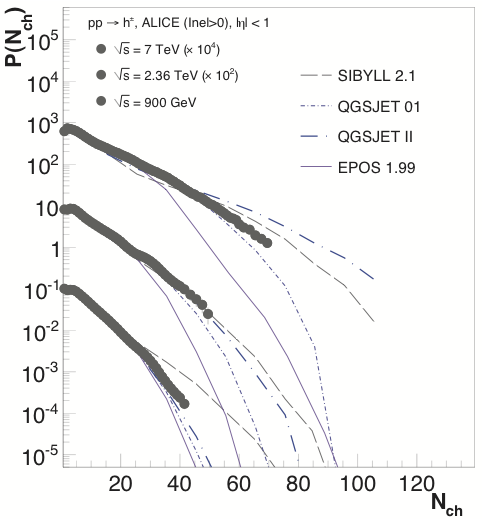}
 \caption{Multiplicity distributions of charged hadrons as measured by ALICE in
	  proton-proton collisions at energies of $\sqrt{s}=0.9$, 2.26, and
          7~TeV compared to predictions of hadronic interaction models
          \iref{Pierog}{1169}.} \label{pierog-mult} 
\end{figure}

The influence of recent LHC results on the interpretation of air-shower data
has been investigated \iref{Pierog}{1169}.  Measurements by ALICE, ATLAS, and
CMS indicate lower values for the proton-proton cross section as predicted by
current hadronic interaction models. If the models are adjusted to the new
measurements one expects a shift of the depth of the shower maximum in air
shower simulations.  This implies that measurements of the shower maximum of
air showers will yield a heavier mass composition of cosmic rays at high(est)
energies.\\
Detailed investigations of pseudo rapidity and multiplicity distributions of
charged secondary particles in proton-proton collisions indicate that the LHC
data are bracketed by predictions of the hadronic interaction models QGSJET-II
and EPOS~1.99.
As an example, the multiplicity distribution of charged hadrons in
proton-proton interactions as measured by ALICE is shown in \fref{pierog-mult}.
Results for different energies are compared to predictions of hadronic
interaction models.
Also distributions measured in forward direction by LHCf are under
investigation. Small deviations are visible with respect to predictions of
hadronic interaction models. However, the influence on the interpretation of
air shower data is expected to be small as compared to the effect of the lower
cross sections.

Studies of hadronic interactions are also under way at the ARGO-YBJ experiment
\iref{De Mitri}{754}.

The author would like to end this section with two remarks:

In the past, various ideas have been discussed as possible origin of the knee
in the energy spectrum of cosmic rays, among them were ideas about new types of
(hadronic) interactions in the atmosphere, see e.g.\ \cite{origin}.  The recent
LHC results exhibit a fair agreement with predictions of hadronic interaction
models using standard physics. Hence, it can be concluded that the knee in the
energy spectrum is not caused by new physics in the atmosphere, it is rather of
astrophysical origin.

The recent results from the LHC (and the Pierre Auger Observatory) indicate
lower values for the proton-air and proton-proton cross sections.  A lower
cross section has already been predicted in 2003 \cite{wq}. In this work the
mean logarithmic mass derived from air shower observations has been
systematically investigated. In particular, a discrepancy has been found
between experiments observing the depth of the shower maximum and experiments
registering the number of secondary particles at ground.  It has been shown
that a smaller proton-proton and consequently also proton-air cross section
reduces the discrepancy in the derived mean logarithmic mass from the different
classes of observables.  The recent results confirm this earlier findings and
this example illustrates how the new LHC measurements directly influence the
interpretation of air shower data.

\Section{HE 3.3 Exotic particle searches}\label{he33}

\subsection{Magnetic monopoles}
Magnetic monopoles are predicted by grand unified theories (GUT)
\iref{Posselt}{734}.  The magnetic charge of a monopole is given by $g =
Ne/2\alpha\approx 68.5~e$, where $\alpha$ is the fine structure constant, and
$e$ the elementary (electric) charge \cite{dirac-monopole}.  The mass of
monopoles is expected to range from $10^8$ to $10^{17}$~GeV for various GUT
models. Because of these large masses magnetic monopoles are generally assumed
to be relics of the early Universe.  Analogous to electric charges, which are
accelerated along electric field lines, magnetic monopoles are accelerated
along magnetic field lines.  During the lifetime of the Universe, relic
monopoles should have encountered enough accelerators to reach kinetic energies
of $\approx10^{14}$~GeV. Thus, monopoles with masses less than
$\approx10^{14}$~GeV should be relativistic.

A water/ice \Cerenkov\ detector is able to detect magnetic monopoles traveling
through the detector at velocities greater than the \Cerenkov\ threshold
($\beta > 0.76$). The radiation emitted by the monopole is proportional to
$(gn)^2$, where $n$ is the index of refraction of the ambient medium. Thus, in
water/ice ($n \approx 1.3$) a monopole will emit almost 8000 times more light
than a single muon of the same velocity.

\begin{figure}[t]
 \includegraphics[width=\columnwidth]{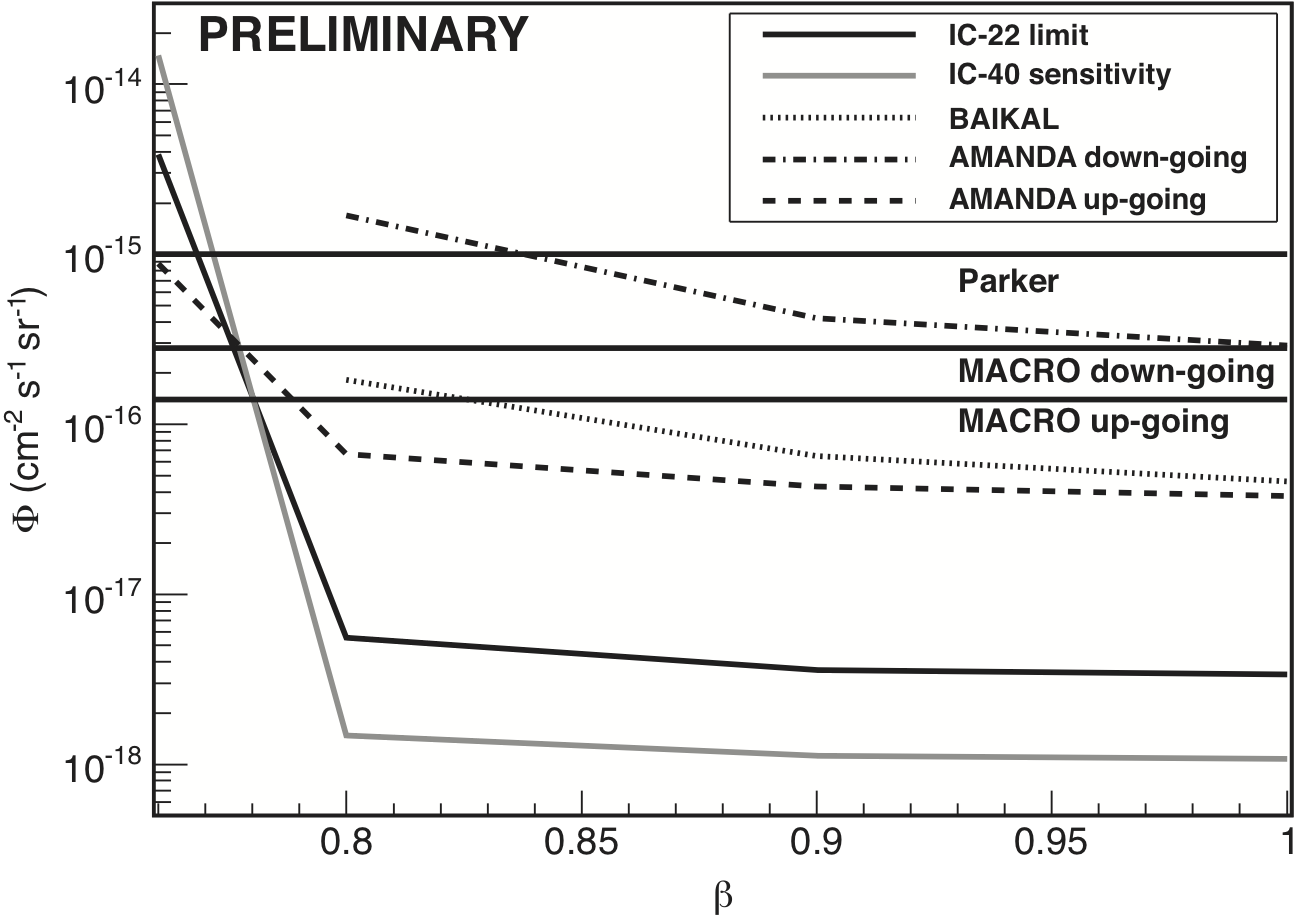}
 \caption{Monopole flux limits set by the 22-string IceCube detector, shown
	  together with previous results \iref{Posselt}{734}. Also shown is the
          expected 40-string sensitivity.}
 \label{icecube-mono}
\end{figure}

Data taken with the 22-string IceCube detector were used to derive new upper
limits on the monopole flux. The limits on the flux of magnetic monopoles with
speeds $\beta> 0.8$ are shown in \fref{icecube-mono}.  They are presently the
most stringent experimental limits to date.  IceCube-40 is expected to reach a
sensitivity of $10^{-18}$~cm$^{-2}$~s$^{-1}$~sr$^{-1}$.  

A similar search has been performed with ANTARES \iref{Picot-Clemente}{695}.
An upper limit has been derived, with values slightly above
$10^{-17}$~cm$^{-2}$~s$^{-1}$~sr$^{-1}$ and, thus, slightly above the limits
set by IceCube.

The imaging atmospheric \Cerenkov\ telescope H.E.S.S. has also been used to
search for monopoles. However, at present, these searches yield less stringent
upper limits as compared to the neutrino telescopes \iref{Spengler}{864}.

\subsection{Antinuclei}
The asymmetry of particles and antiparticles in the Universe is one of the
fundamental questions in cosmology \iref{Sasaki}{1230}.  While antiprotons have
been found in cosmic rays, so far no antiparticles with $|Z|\ge2$ have been
detected.  Many cosmologists consider that this asymmetry was caused by the
symmetry breaking between particles and antiparticles just after the Big Bang,
with cosmological antiparticles vanishing at an early stage of the Universe.
However, the existence of $|Z|\ge2$ antiparticles is not excluded by theory.
There might be remnant antiparticle domains from the Big Bang.

Therefore, the BESS Collaboration has conducted an extensive search  for
antinuclei in cosmic rays over the last 20 years \iref{Sasaki}{1230}
\iref{Yoshimura}{1259}. The heart of BESS is a magnetic spectrometer to
identify the sign of the throughgoing particles, and, thus, identify
antiparticles.  The set-up is complemented by time-of-flight counters and a
silica aerogel \Cerenkov\ counter.  Various experimental configurations have
been developed and flown on stratospheric balloons in the last two decades.
Results from the latest version of the instrument, BESS-Polar II, have been
reported.

\begin{figure}[t]
 \includegraphics[width=\columnwidth]{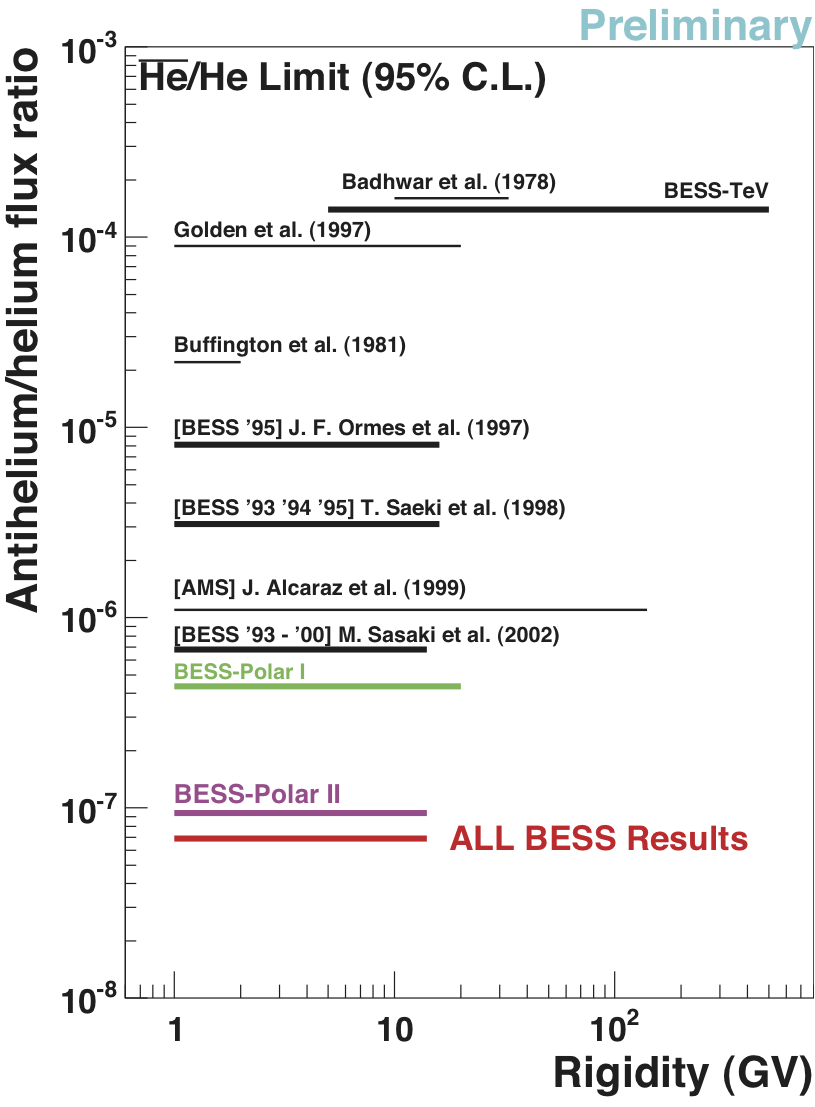}
 \caption{Upper limits on the ratio of antihelium to helium nuclei in cosmic
          rays as reported by the BESS experiment \iref{Sasaki}{1230}.}
 \label{antihe}
\end{figure}

The upper limits on the antihelium-to-helium ratio are depicted in
\fref{antihe}. The sensitivity has improved over the last 20 years by a factor
of 1000 -- a quite impressive number. The recent results indicate that there
is less than 1 antihelium nucleus per $10^7$ helium nuclei in the Universe.

Searches for antideuterons are under way \iref{Yoshimura}{1259}. No
antideuteron has been found so far in cosmic rays.  Upper limits are expected
to be published soon.

\subsection{Simultaneous air showers}
The coincident detection of air showers from the same direction at the same
time would be an experimental hint towards interesting physical processes.
Among them are speculations about strangelets from cosmological origin or the
break-up of cosmic-ray nuclei in the heliosphere.

A search for simultaneous showers was conducted at CERN for strangelets whose
breakup in space may result in the arrival of a large number of time and
angle-coincident air showers spread over a large area \iref{Tonwar}{1295}.  Two
shower detector arrays were used, located on the surface at points P2 and P4 at
the LHC, which are separated horizontally by about 8~km. The arrival time of
showers was recorded at each array with 100 ns accuracy, and the spatial
arrival angle of showers was determined to an accuracy of about $3^\circ$.
More than $10^7$ showers were collected between 2004 and 2006.\\
No pairs of showers were detected arriving within $30~\mu$s in time and within
$5^\circ$ in spatial angle at the two stations.  This yields a 90\% c.l.\ upper
limit of $5.1 \cdot 10^{-20}$~cm$^{-2}$~sr$^{-1}$~s$^{-1}$ on the flux of
strangelets.

A search was made for the interaction between cosmic rays and ions in the
heliosphere using data collected by the HiRes-1 detector of the High Resolution
Fly's Eye \iref{Rodriguez}{1304}.  Such an interaction could produce a jet
containing neutral pions which can lead to an observation of an  unique
signature of parallel, simultaneous photon showers. The data were searched in a
time window of $100~\mu$s. No coincident showers have been found.

\Section{HE 3.4 Direct and indirect dark matter searches}\label{he34}

According the current standard model of cosmology, almost 25\% of the energy
budget of the Universe is an unknown, invisible mass component, the dark
matter.  Despite many efforts over the last years, dark matter is still one of
the greatest mysteries of modern physics.  The dark-matter particle could be a
weakly interacting massive particle (WIMP). It is assumed that WIMPs were
thermally produced in the early Universe, that they are stable, and of a
non-baryonic nature \cite{Bertone:2004pz}.  Through self-annihilation, WIMPs
should produce Standard Model particles, and some of the byproducts, like
photons, hadrons and leptons, might be observable at Earth. Supersymmetry
(SUSY) provides a natural candidate for the dark-matter WIMP -- the lightest
SUSY particle neutralino.

Given the natural mass range for the SUSY WIMPs (between a few GeV and a few
TeV) \cite{Bertone:2010zz}, it may be possible to find, via very energetic
photons, some signatures for dark-matter annihilation in the energy range of
imaging atmospheric \Cerenkov\ telescopes \iref{Aleksic}{331}.  The typical
annihilation gamma-ray spectrum is predicted to be continuous and featureless,
due to the gammas mainly being produced from pion decays and final-state
radiation of charged particles. Nevertheless, some distinctive spectral
features could be present, like the line emission (from WIMP annihilation into
a pair of gammas or a gamma and a $Z$-boson), a cut-off or spectral hardening
due to internal bremsstrahlung, with all of these dependent on the mass of the
dark-matter particle.

The basic search strategy for indirect dark-matter searches is to look for
high-energy gamma rays and neutrinos from the annihilation of dark-matter
particles.  Given that the annihilation rate is proportional to the dark-matter
density squared, the best places to look for the WIMPs are astronomical regions
with a large concentration of dark matter. Several recent activities have been
discussed, such as looking for annihilation products in 
galaxy clusters,
dwarf spheroidal galaxies,
globular clusters,
the Milky Way halo, or the Sun.

\subsection{Galaxy clusters}
Clusters of galaxies, such as the Virgo Cluster, host enormous quantities of
dark matter, making them prime targets for efforts in indirect dark matter
search.  A multi-wavelength spectral energy distribution for the central
radio galaxy in the Virgo Cluster, M87, has been constructed, using a
state-of-the-art numerical synchrotron self Compton approach. Fitting recent
Chandra, FERMI-LAT, and \Cerenkov\ telescope observations, yields a best fit
value for the neutralino mass of $m_\chi=3.4$~TeV \iref{Saxena}{204}.

\subsection{Dwarf spheroidal galaxies}
Dwarf spheroidals, the satellite galaxies of the Milky Way, have been
identified as one of the most suitable candidates for dark-matter searches: the
ratio between the mass inferred from gravitational effects and the mass
inferred from luminosity $(M/L)$ can reach values from 100 up to few $1000~
M_\odot /L_\odot$ for these objects.  They are relatively close to the observer
and they are expected to be free of astrophysical objects whose gamma-ray
emission might ’hide’ the dark-matter signal \iref{Aleksic}{331}.

Until several years ago, only nine dwarf spheroidal galaxies were discovered
orbiting the Milky Way. These, so-called ‘classical‘ dwarfs were joined by 11
new, ultra-faint objects thanks to the discoveries of the Sloan Digital Sky
Survey (SDSS), starting in 2005. Among these objects several have been
identified as excellent candidates for indirect dark-matter searches
\iref{Aleksic}{331}.

Segue 1 is a ultra-faint object discovered in 2007 by the SDSS and located
23~kpc from the Sun.  It is the most dark-matter dominated object known so far,
with $M/L \approx 1320-3400~M_\odot/L_\odot$.

\begin{figure}[t]
 \includegraphics[width=\columnwidth]{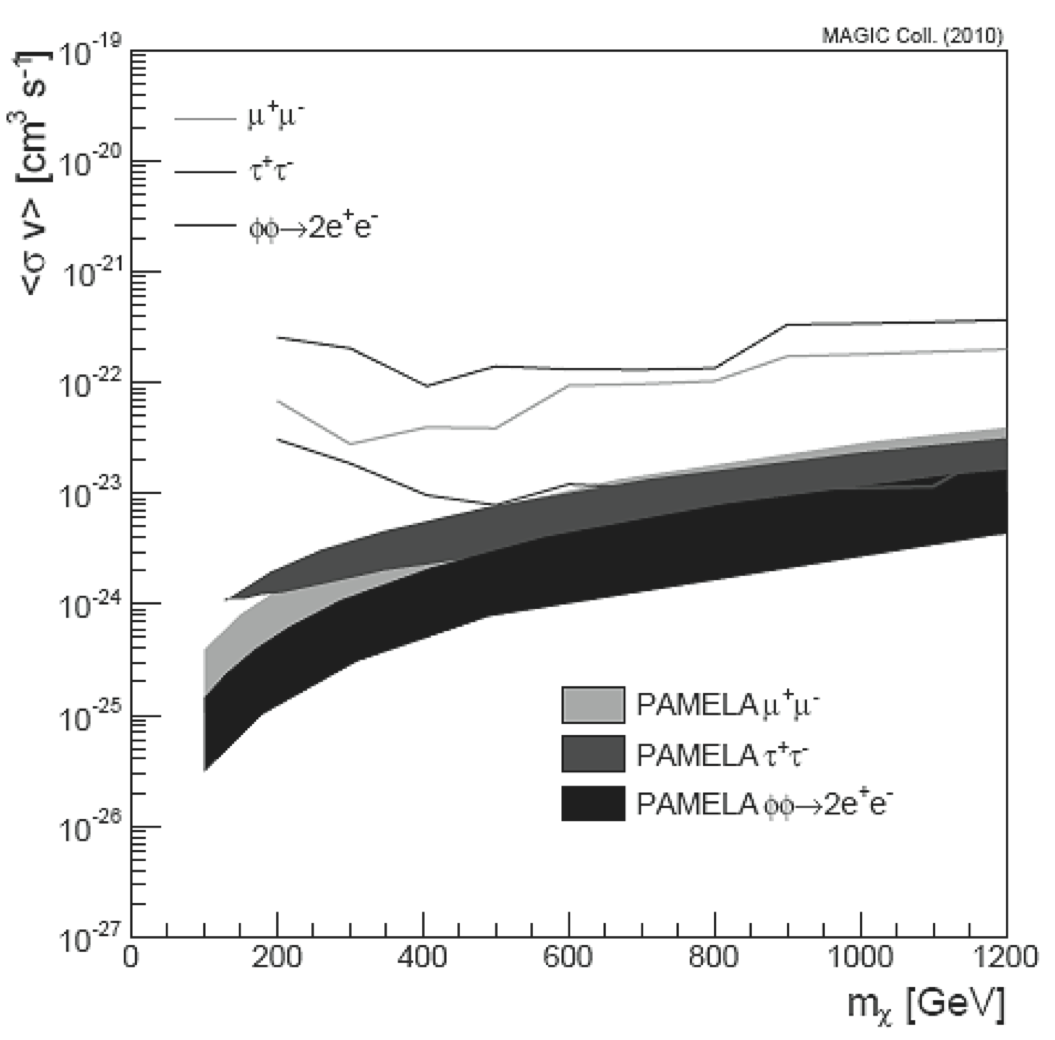}
 \caption{Upper limits on particular annihilation channels as obtained by
	  MAGIC and corresponding regions that provide a good fit to PAMELA
          data \iref{Aleksic}{331}.}
 \label{magic-dm}
\end{figure}

Segue 1 has been observed with the MAGIC \Cerenkov\ telescope for 29.4 hours
\iref{Aleksic}{331}.  No significant gamma-ray signal was found above the
background.  The data have been used to probe dark-matter models that  try to
explain the PAMELA data, see \fref{magic-dm}.  Current results are probing the
PAMELA-preferred region for the case of dark-matter annihilation into $\tau^+ +
\tau^-$.

Segue 1 has also been observed by VERITAS for almost 48 hours
\iref{Vivier}{919}.  For the $W^+ W^-$ annihilation channel, a 95\% c.l.\ upper
limit on the velocity-weighted annihilation cross section $\langle\sigma
v\rangle$ has been obtained at the level of $8 \cdot 10^{-24}$~cm$^3$~s$^{-1}$
around 1 TeV.

13 dwarf spheroidal galaxies (also among them Segue~1) have been observed by
IceCube-59 to search for dark matter \iref{L\"unemann}{1024}.  Sensitivities to
the velocity-weighted dark matter self-annihilation cross section are better
than $10^{-20}$~cm$^3$~s$^{-1}$, for WIMP masses in a range from 300 GeV to
several TeV. These searches are complementary to gamma-ray observations in most
channels and have the advantage that IceCube data of these sources are
collected continuously and extend to higher WIMP masses.

The Sculptor and Carina dwarf spheroidal galaxies were observed with H.E.S.S.
for 11.8 and 14.8 hours, respectively \iref{Viana}{1036}.  No gamma-ray signal
was detected at the nominal positions of these galaxies above 220~GeV and
320~GeV, respectively.  Constraints on the dark matter velocity-weighted
annihilation cross section for both, the Sculptor and Carina dwarf galaxies
range from $\langle\sigma v\rangle \approx 10^{-21}$ down to
$10^{-22}$~cm$^3$~s$^{-1}$, depending on the dark-matter halo model.

\subsection{Globular clusters}
Several Galactic globular clusters have been observed with imaging atmospheric
\Cerenkov\ telescopes.  Globular clusters are dense stellar systems with an age
of about 10 Gyr, found in halos of galaxies, with typical masses between $10^4$
and a few $10^6~M_\odot$.  They are potential targets for indirect dark-matter
searches.

\begin{figure}[t]
 \includegraphics[width=\columnwidth]{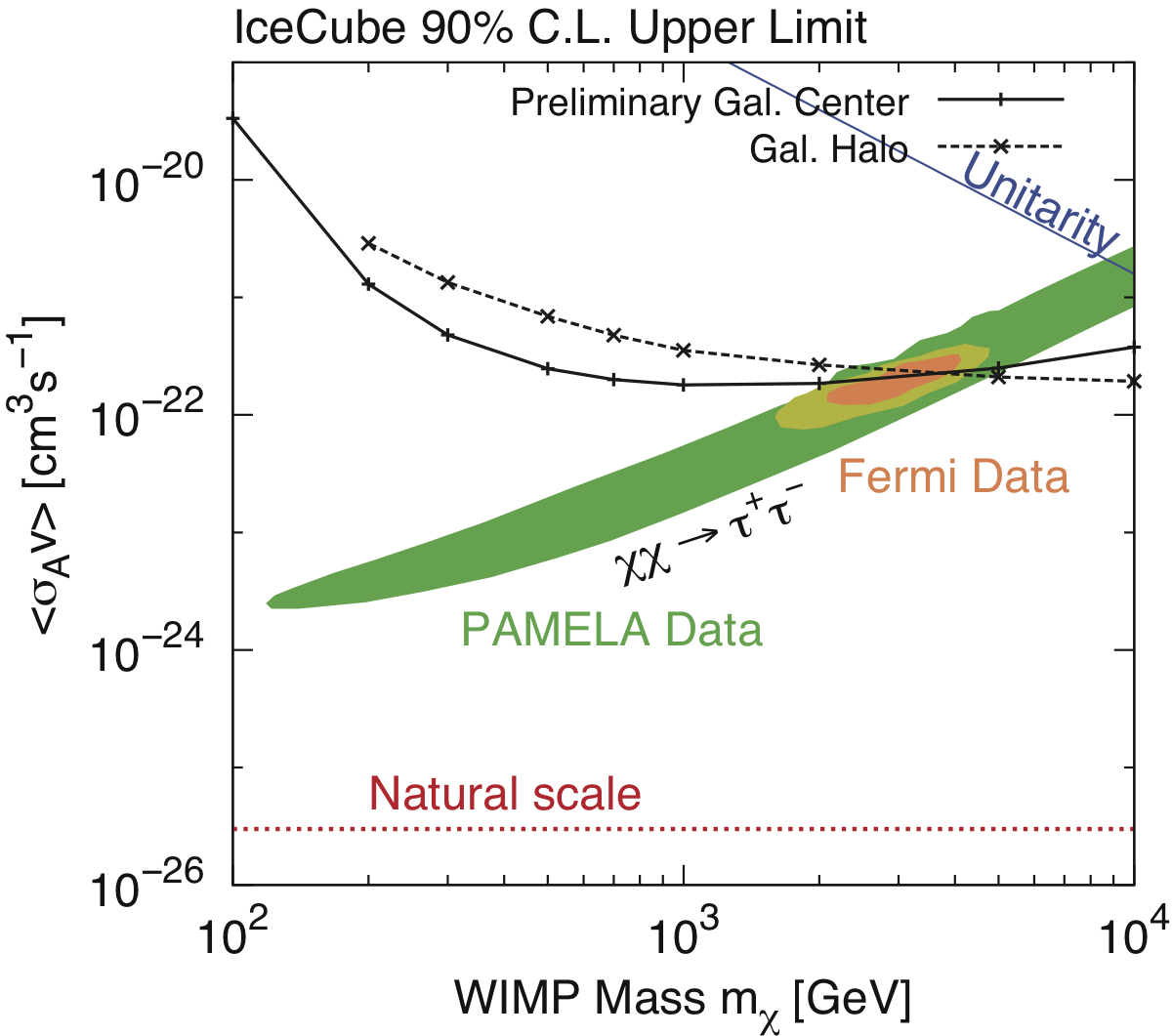}
 \caption{Limit on the velocity-weighted WIMP cross section $\langle \sigma
	  v\rangle$ at 90\% c.l.\ as function of the WIMP mass as obtained from
	  observations of the Galactic Center with IceCube \iref{Rott}{1187}.
	  The preferred range obtained from gamma-ray observations with PAMELA
          and Fermi is indicated as well.}
 \label{icecube-dmmilkyway}
\end{figure}

The globular clusters NGC 6388 and M15 have been observed with H.E.S.S.\ for a
lifetime of 27.2 and 15.2 hr, respectively \iref{Moulin}{1121}.  No gamma-ray
signal is found at the nominal target positions.  95\% c.l.\ exclusion limits
on the dark matter velocity-weighted annihilation cross section
$\langle\sigma v\rangle$ are derived for these dark matter halos. In the TeV
range, these limits reach the $10^{-25}$~cm$^3$~s$^{-1}$ level and a few
$10^{-24}$~cm$^3$~s$^{-1}$ for NGC 6388 and M15, respectively.

The Fornax galaxy cluster has been observed with H.E.S.S.\ for a total of 11
hours of lifetime.  For a dark matter particle mass of 1~TeV, the exclusion
limits reach values of $\langle\sigma v\rangle < 10^{-21}$~cm$^3$~s$^{-1}$,
depending on the dark-matter model and halo properties \iref{Viana}{1038}.

\subsection{Milky Way halo}
The central region of the dark-matter halo in the Milky Way is also a promising
target for a search for a particle dark matter self-annihilation signal.  With
the IceCube neutrino telescope a search for WIMP dark matter accumulated in the
Galactic halo and the Galactic center has been conducted \iref{Rott}{1187}.
Limits on the dark matter velocity-weighted annihilation cross section
$\langle\sigma v\rangle$ at a level of $10^{-22}$ to $10^{-23}$~cm$^3$~$s^{-1}$
have been achieved.  The values obtained are shown in \fref{icecube-dmmilkyway}
as function of WIMP mass together with recent results from the PAMELA and Fermi
gamma-ray space missions. The various experimental values nicely overlap for
WIMP masses in the region $2-3$~TeV.

Corresponding searches with the H.E.S.S.  imaging atmospheric \Cerenkov\
telescope are under way \iref{Spengler}{862}.

\subsection{Sun}
\begin{figure}[t]
 \includegraphics[width=\columnwidth]{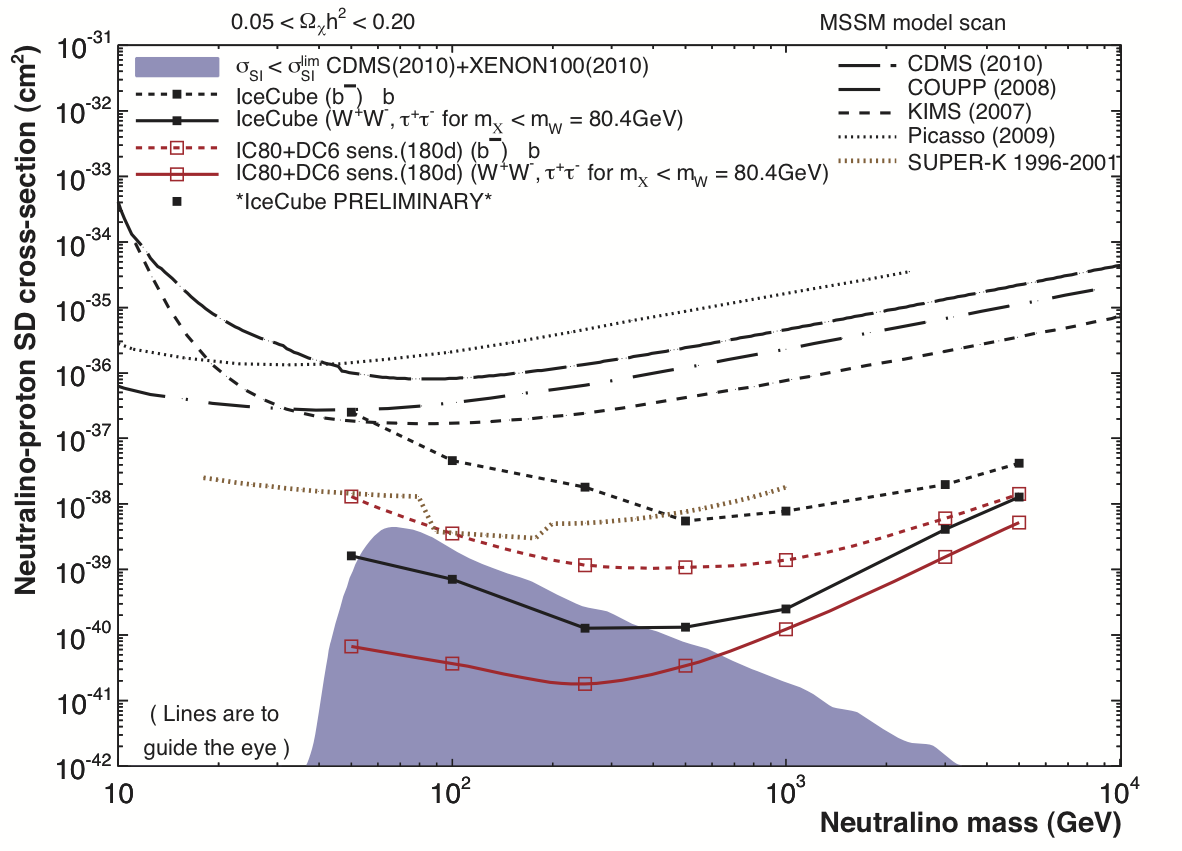}
 \caption{Sensitivity on the spin-dependent neutralino-proton cross section at
          90\% c.l.\ as function of the neutralino mass as obtained by IceCube
          \iref{Danninger}{292}.}
 \label{icecube-dmsun}
\end{figure}

Another possibility to look for dark-matter signatures is the Sun, where dark
matter could be gravitationally trapped.  Dark matter could be indirectly
detected through the observation of neutrinos produced as part of the
self-annihilation process.

The IceCube neutrino telescope with the DeepCore sub-array is used to
search for dark matter \iref{Danninger}{292}.  This will lead to stringent
constraints on WIMP-proton cross sections.  The expected sensitivity of IceCube
is illustrated in \fref{icecube-dmsun}.  For neutralino masses in the 100~GeV
range, limits on the proton-neutralino cross section as low as
$10^{-41}$~cm$^{-2}$ are expected.  Recent limits obtained with AMANDA and
IceCube 40 are discussed in \iref{Engdegard}{327}.

ANTARES reports limits for the spin-dependent neutralino cross section, with
sensitivities similar to the preliminary IceCube results \iref{Lambard}{202}.

\Section{Acknowledgments}
The author would like to thank the organizers of the conference for their great
hospitality in Beijing.

I'm grateful to John Kelley for critically reading the manuscript.


\end{document}